\documentclass[11pt,fleqn]{article}
\usepackage[top=0.8in,bottom=0.7in,left=0.2in,right=0.2in,headheight=14pt,headsep=10pt,columnsep=14pt]{geometry}
\usepackage{amsmath,amssymb}
\usepackage{bm}
\usepackage{graphicx}
\usepackage{titlesec}
\usepackage{algorithm}
\usepackage{algorithmic}
\usepackage{etoolbox}

\AtBeginEnvironment{equation}{\noindent}
\AtBeginEnvironment{equation*}{\noindent}
\usepackage{fancyhdr}
\usepackage[numbers,square]{natbib}

\usepackage{listings}
\usepackage[svgnames]{xcolor}
\usepackage{hyperref}
\hypersetup{
    colorlinks=true,
    linkcolor=blue,
    citecolor=blue,
    urlcolor=blue,
    filecolor=blue
}
\usepackage{cleveref}
\usepackage{caption}
\usepackage{footmisc}
\usepackage{authblk}
\usepackage{multicol}

\titlespacing*{\section}{0pt}{20pt}{10pt}
\titlespacing*{\subsection}{0pt}{16pt}{6pt}
\titlespacing*{\subsubsection}{0pt}{8pt}{3pt}

\titleformat*{\section}{\large\bfseries}
\titleformat*{\subsection}{\normalsize\bfseries}
\titleformat*{\subsubsection}{\normalsize\bfseries}

\fancypagestyle{plain}{%
    \fancyhf{}%
    \fancyhead[L]{\hspace*{0.3in}R. Meng and X. Dong (2026)}%
    \fancyhead[R]{\thepage\hspace*{0.3in}}%
}

\newcommand{\grad}{\nabla}
\newcommand{\sech}{\operatorname{sech}}
\newcommand{\etal}{et~al.}

\DeclareMathSizes{10}{9.5}{7}{5}

\newenvironment{keywords}{%
  \begin{quote}\noindent\textbf{Keywords:}~%
}{\end{quote}}

\lstdefinelanguage{Julia}%
  {morekeywords={abstract,begin,break,case,catch,const,continue,do,else,elseif,%
      end,export,false,for,function,immutable,import,importall,if,in,%
      macro,module,otherwise,quote,return,switch,true,try,type,typealias,%
      using,while},%
   sensitive=true,%
   alsoother={\$},%
   morecomment=[l]\#,%
   morecomment=[n]{\#=}{=\#},%
   morestring=[s]{"}{"},%
   morestring=[m]{'}{'},%
}[keywords,comments,strings]%

\definecolor{codegray}{rgb}{0.5,0.5,0.5}

\begin{document}

\title{WaveDM.jl: An Adaptable Simulation Framework for Dynamics of Baryonic and Wave Dark Matter on Galaxy Scales}

\author[a,b]{Run-Yu Meng}
\author[a,b]{Xiao-Bo Dong\thanks{Corresponding authors. Email: xbdong@ynao.ac.cn; mry1234@mail.ustc.edu.cn}}

\affil[a]{Yunnan Observatories, Chinese Academy of Sciences, Kunming, Yunnan 650216, China}
\affil[b]{University of Chinese Academy of Sciences, Beijing 100049, China}

\date{\today}

\maketitle

\abstract{
    We present \texttt{WaveDM.jl}, an open-source Julia package for high-performance simulations of wave dark matter dynamics on galaxy scales,
    with a design philosophy centered on extensibility and adaptability.
    The code solves the time-dependent Schr\"odinger--Poisson equation (SPE) using a pseudo-spectral Fourier method.
    The spectral solver is tightly integrated with $N$-body gravitational force solvers,
    enabling simultaneous evolution of wave dark matter and baryonic components in galaxy-scale simulations.
    \texttt{WaveDM.jl} unifies shared-memory, distributed-memory, and GPU execution within a multi-level parallelization framework,
    enabling the same computational workflow to scale from single-node to multi-node computing environments without requiring major code changes.
    To further facilitate user-friendly galaxy-scale simulations,
    the package provides a dedicated toolbox that integrates
    flexible initial condition generators, trajectory lookback, tidal force calculations, and real-time visualization. 
    Beyond astrophysical applications,
    the code's modular architecture and general nonlinear Schr\"odinger framework
    enable cross-disciplinary studies such as nonlinear optics and cold-atom physics.
    The code is an open source and available on \url{https://github.com/JuliaAstroSim/WaveDM.jl}.
}


\begin{multicols}{2}

\section{Introduction}\label{sec:intro}

Fuzzy cold dark matter (also called wave CDM) represents a compelling variant of the collisionless CDM paradigm \citep{hu2000fuzzy,hui2021wavedark}.
Originally proposed as a possible way to address the small-scale challenges of conventional particle CDM \citep{hu2000fuzzy}, 
this model describes dark matter as ultralight bosons with masses typically in the range 
$m_a \sim 10^{-23}$--$10^{-20}\,{\rm eV}$. 
For such small particle masses, the de~Broglie wavelength,
$\lambda_{\rm dB} \equiv {h}/{(m_a v)}$,
can reach astrophysical scales in galactic halos. 
For example, for $m_a = 10^{-22}\,{\rm eV}$ and a characteristic velocity inside a galactic halo 
$v=100\,{\rm km\,s^{-1}}$, one obtains $\lambda_{\rm dB} \approx 1.2\,{\rm kpc}$,
which is comparable to the characteristic size of the main components of the galaxy such as the bulge. 
As a result, wave CDM exhibits phenomena absent in conventional particle CDM, 
including solitonic cores presumably supported by quantum pressure, 
granules mainly due to wave interference, wave-like halo fluctuations, 
and the suppression of structure below the de~Broglie scale. 
Potentially, these features can affect central densities of dwarf galaxies, 
the abundance of low-mass subhalos, and the dynamical evolution of baryonic tracers in galactic potentials (see \cite{hui2021wavedark} for a review). 
Accurately modeling these wave phenomena on galactic scales requires numerical tools capable of 
simultaneously capturing the global structure of DM halos 
and resolving those wave patterns on spatial scales much smaller than their local de~Broglie wavelengths \citep{li2019numerical,schive2026fuzzydark}.

Current wave CDM simulations primarily employ two complementary approaches:
solving the wave evolution equation of the self-gravitating system on grids (namely the Schr\"odinger--Poisson equation (SPE); 
see \S \ref{sec:methods_theoretical} for details),
or reformulating the system via the equivalent formulation (Madelung equations) into a fluid-dynamics treatment.
SPE solvers provide spectral accuracy for capturing wave interference patterns and topological defects.
Madelung-based methods offer computational efficiency when simulating large volumes
but face challenges in regions of destructive interference
where the fluid description breaks down.
See \cite{schive2026fuzzydark} for a comprehensive review of these numerical methodologies.

Existing wave CDM simulation packages have demonstrated diverse capabilities, 
yet a unified framework that combines accessibility, performance, and specialized tools for galaxy-scale studies remains absent.
For example, \texttt{PyUltraLight} \citep{edwards2018pyultralight} offers an accessible pseudo-spectral implementation
but lacks parallelization capabilities for high-resolution simulations.
\texttt{UltraDark.jl} \citep{musoke2024ultradarkjl} and \texttt{GAMER} \citep{schive2014cosmicstructure, kunkel2025hybrid}
employ MPI (Message Passing Interface) parallelization \citep{gropp1999usingmpi} for high performance,
though adapting them to specific scientific questions typically involves significant customization effort.
Additionally,
these codes lack the specialized tools essential for galaxy-scale simulations.

To address these limitations,
we present \texttt{WaveDM.jl} --- a high-performance Julia package for wave dark matter research on galaxy scales.%
\footnote[1]{The code can also be applied to simulations of non-relativistic wave warm dark matter
\citep{liu2025warmfuzzy}.}
Julia is a modern programming language that combines high performance with ease of use, 
specifically designed for scientific computing and data analysis. 
Leveraging these advantages,  
\texttt{WaveDM.jl} integrates an SPE solver with $N$-body gravitational methods within a multi-level parallel framework, 
built on a modular architecture that facilitates adaptability and extensibility. 
This design provides high-level interfaces that allow users to focus on astrophysical analysis 
rather than managing numerical implementations. 
Users interact with the package by writing short Julia scripts (similar to Python or Jupyter notebooks) 
to configure simulations and analyze results, without modifying the source code. 
Specifically, \texttt{WaveDM.jl} offers three key capabilities: 
simultaneous evolution of wave dark matter and baryonic components, 
multi-level parallel execution across diverse computing environments, 
and a dedicated galaxy-simulation toolbox.

The remainder of this paper is organized as follows.
Sect.~\ref{sec:methods_theoretical} establishes the theoretical framework of wave CDM simulations.
Sect.~\ref{sec:methods_numerical} describes the numerical methods.
Sect.~\ref{sec:toolbox} demonstrates the code's capabilities through representative galaxy simulations.
Sect.~\ref{sect:tests} presents comprehensive convergence tests validating numerical accuracy.
Sect.~\ref{sect:discussion} discusses our design philosophy, 
addresses limitations and future development plans, 
and explores cross-disciplinary applications.
Finally, Sect.~\ref{sect:summary} summarizes the key features and applications of \texttt{WaveDM.jl}.

\section{Theoretical Framework} \label{sec:methods_theoretical}

\subsection{Basic physics of wave CDM halos} \label{sec:methods_C-field}

Being composed of ultralight bosons,
the most fundamental characteristic of wave CDM in the context of galactic-scale dark matter halos 
is this: 
the extremely high critical temperature for Bose--Einstein condensation, $T_c \sim 10^{35}\,{\rm K}$
\citep{chavanis2019predictive}.
By comparison, the cosmic microwave background temperature is approximately $3\,{\rm K}$, 
and the effective temperature associated with the kinetic energy of the bosons is only
$T_{\rm eff} \sim 10^{-25}\,{\rm K}$ for the fiducial boson mass $m_a \sim 10^{-22}\,{\rm eV}$.
Thus, relative to $T_c$, the system is effectively in the zero-temperature regime.

In this regime, the dynamics of a highly Bose-degenerate bosonic system can be well described
by a single complex-valued macroscopic wavefunction $\psi(\mathbf r,t)$, 
which plays the role of the order parameter.
Once the system reaches a (quasi-)equilibrium state, 
almost all bosons are expected to reside in the condensate.
The ground-state component of the condensate is often referred to as the ``soliton'' in the wave-CDM literature, 
while the excited component of the condensate
\footnote[2]{See \cite{dong2025promise} for the astrophysical implications,
particularly concerning the recent puzzles from JWST.}
carries collective excitations, 
or called quasiparticles, such as phonons, vortices, and various spatially localized modes.
The evolution of the order parameter is governed by the well-known Gross--Pitaevskii equation (GPE).
See \cite{pitaevskii2016boseeinstein} further for the theoretical background.

It is important to note that, in this $T\,\rightarrow\,0$ regime, 
the applicability of a macroscopic wave function is not limited to systems 
that have already established global phase coherence. 
Even if the bosonic system is initially far from quasi-equilibrium, 
and even in the absence of a fully developed Bose--Einstein condensate, 
its evolution can still be described by the classical-field GPE 
under the conditions relevant for wave-CDM halos \citep{kagan1997evolutioncorrelation}. 
This is because the system is highly Bose degenerate: 
the de~Broglie wavelengths of the bosons overlap strongly, 
and many low-energy single-particle modes are highly occupied. 
In this regime, $\psi(\mathbf r,t)$ should be interpreted as a classical matter-wave field 
representing highly occupied bosonic modes; 
see \cite{kagan1997evolutioncorrelation} for a detailed discussion.

\subsection{The Schr\"odinger--Poisson system} \label{sec:methods_SPE}

The governing equation of $\psi(\mathbf r,t)$, namely GPE, is 
coupled self-consistently with the Poisson equation of gravity in the case of DM halos:
\begin{equation} \label{eqn:GPE}
    i\hbar \frac{\partial\psi}{\partial t}(\mathbf{r},t) = -\frac{\hbar^2}{2m}\nabla^2 \psi(\mathbf{r},t) + m\Phi_{\mathrm{total}}(\mathbf{r},t)\psi(\mathbf{r},t),
\end{equation}

\begin{equation}
    \label{eqn:Poisson}
    \nabla^2 \Phi_{\mathrm{total}}(\mathbf{r},t) = 4\pi G\left[\rho_{\mathrm{DM}}(\mathbf{r},t) + \rho_{\mathrm{baryon}}(\mathbf{r},t)\right],
\end{equation}
where $\rho_{\mathrm{DM}} = m |\psi(\mathbf{r},t)|^2$ represents the mass density of the wave dark matter,
$m$ denotes the boson mass,
and $\Phi_{\mathrm{total}}$ incorporates gravitational contributions from both the wave CDM and baryonic components.

The Madelung transformation provides a hydrodynamic representation of the SPE.
By expressing $\psi$ in polar form as:
\begin{equation}
    \label{eqn:modulus-arg}
    \psi(\mathbf{r},t) = \sqrt{\frac{\rho(\mathbf{r},t)}{m}} \, e^{iS(\mathbf{r},t)/\hbar},
\end{equation}
where $S(\mathbf{r},t)$ is the phase function,
we derive the velocity field:
\begin{equation}
    \label{eqn:fluid_velocity_field}
    \mathbf{u}(\mathbf{r},t) = \frac{1}{m}\grad S(\mathbf{r},t).
\end{equation}
This velocity field is irrotational, satisfying $\grad \times \mathbf{u} = 0$,
as expected for a quantum fluid derived from a scalar wave function.

Substituting Eq.~\ref{eqn:modulus-arg} into the Schr\"odinger equation Eq.~\ref{eqn:GPE}
and separating real and imaginary parts yield the continuity equation
\begin{equation}
    \label{eqn:Mad_continuity}
    \frac{\partial\rho}{\partial t} + \nabla\cdot (\rho \mathbf{u}) = 0,
\end{equation}
and the quantum version of the generalized Bernoulli equation for unsteady flows
\begin{equation}
    \label{eqn:Mad_Bernoulli}
    \frac{\partial S}{\partial t} + \frac{m}{2}\mathbf{u}^2 + m\Phi_{\mathrm{total}} + Q = 0,
\end{equation}
where $Q$ is the Bohm potential (quantum potential), given by:
\begin{equation}
    \label{eqn:Potential_Q}
    Q = -\frac{\hbar^2}{2m}\frac{\nabla^2 \sqrt{\rho}}{\sqrt{\rho}} = -\frac{\hbar^2}{4m}\left\lbrack \frac{\nabla^2\rho}{\rho} - \frac{1}{2}\frac{(\nabla\rho)^2}{\rho^2} \right\rbrack .
\end{equation}
The Bohm potential accounts for quantum effects, including the Heisenberg uncertainty principle.

Taking the gradient of Eq.~\ref{eqn:Mad_Bernoulli} and using the vector identity
$(\mathbf{u}\cdot \nabla)\mathbf{u} = \nabla \left(\frac{\mathbf{u}^2}{2}\right) - \mathbf{u}\times (\nabla\times \mathbf{u})$, 
which reduces to $(\mathbf{u}\cdot \nabla)\mathbf{u} = \nabla \left(\frac{\mathbf{u}^2}{2}\right)$ for irrotational flows,
yield the quantum Euler equation for wave CDM:

\begin{equation}
    \label{eqn:Mad_Euler}
    \frac{\partial \mathbf{u}}{\partial t} + (\mathbf{u}\cdot \nabla)\mathbf{u} = -\grad\Phi_{\mathrm{total}} - \frac{1}{m} \grad Q ~.
\end{equation}

For computational convenience, we adopt the length, time, and mass scales consistent with \cite{edwards2018pyultralight} and \cite{glennon2021modifying}:
\begin{equation}
    \label{eqn:code_units}
    \begin{aligned}
        \mathcal{L} &= \left(\frac{8\pi\hbar^{2}}{3m^{2}H_{0}^{2}\Omega_{m_{0}}}\right)^{\frac{1}{4}}\approx 121\left(\frac{10^{-23} \mathrm{eV}}{m}\right)^{\frac{1}{2}}\mathrm{kpc} ~,\\
        \mathcal{T} &= \left(\frac{8\pi}{3H_{0}^{2}\Omega_{m_{0}}}\right)^{\frac{1}{2}}\approx 75.5 \mathrm{Gyr} ~,\\
        \mathcal{M} &= \frac{1}{G}\left(\frac{8\pi}{3H_{0}^{2}\Omega_{m_{0}}}\right)^{-\frac{1}{4}}\left(\frac{\hbar}{m}\right)^{\frac{3}{2}}\approx 7\times10^{7}\left(\frac{10^{-23} \mathrm{eV}}{m}\right)^{\frac{3}{2}}\mathrm{M}_{\odot} ~.
    \end{aligned}
\end{equation}
Here, $H_0$ is the Hubble constant and $\Omega_{m_0}$ is the matter density parameter.
In dimensionless units, the SPE takes the form (absorbing dimensional quantities for notational convenience):
\begin{equation} \label{eq:SPE_dimensionless}
    \begin{aligned}
        i \frac{\partial\psi}{\partial t} &=-\frac{1}{2} \nabla^{2} \psi + \Phi_{\mathrm{total}} \psi,\\
        \nabla^{2} \Phi_{\mathrm{total}} &= 4\pi (\rho_{\mathrm{DM}} + \rho_{\mathrm{baryon}}).
    \end{aligned}
\end{equation}
In this dimensionless form,
$\psi = \sqrt{\rho} e^{i \theta}$,
and the velocity field is $\mathbf{v} = \nabla \theta$.

In the wave DM literature, 
interactions between bosons beyond gravity are sometimes added to the right-hand side of the Schr\"odinger equation.
This so-called self-interaction term typically takes the form $\kappa |\psi|^2 \psi$, 
with $\kappa = 4\pi\hbar a_s / (\mathcal{T} m^2)$, 
where $a_s$ is the $s$-wave scattering length 
(positive and negative signs correspond to repulsive and attractive interactions, respectively). 
\texttt{WaveDM.jl} supports non-interacting ($\kappa = 0$), repulsive ($\kappa > 0$), and attractive ($\kappa < 0$) scenarios. 
In fact, the code accommodates all kinds of interactions between particles,
both local and nonlocal (see Sect.~\ref{sec:methods_SPE_general} for details). 
In the simulations and case studies presented in this paper, we yet focus on the non-interacting case.

\subsection{Quantum pressure, quantum potential and gradient energy}
\label{sec:methods_SPE_Egradient}

The density-weighted integral of the quantum potential $Q$
will be used in the overall-energy diagnostics below (see Sect.~\ref{sec:methods_virialization}).
In the literature, several terms are used for this volume-integrated physical quantity, such as
\emph{quantum energy} \citep{hui2017ultralight}, \emph{gradient energy} ($K_\rho$, \cite{mocz2017galaxy})
and \emph{effective thermal energy} \citep{liao2025decipheringsolitonhalo}, 
without a clear explanation of its relationship with quantum pressure and quantum potential.
Below we present a mathematical derivation and reveal the connections among these quantities.

Multiplying Eq.~\ref{eqn:Potential_Q} by $\rho$, integrating over space, and applying Green's first identity give
\begin{equation}
    \label{eqn:Q_integration}
    \begin{aligned}
        \mathcal{Q} & \equiv \int \rho Q \,\mathrm{d}^3r \\
        &= -\frac{\hbar^2}{2m} \int \sqrt{\rho}\,\nabla^2\sqrt{\rho} \,\mathrm{d}^3r \\
        &= - \frac{\hbar^2}{2m} \oint_{\partial V} \sqrt{\rho}\,\nabla\sqrt{\rho}\cdot\mathrm{d}\mathbf{S} 
           \, + \, \frac{\hbar^2}{2m} \int |\nabla\sqrt{\rho}|^2 \,\mathrm{d}^3r  \, .
    \end{aligned}
\end{equation}
The above surface term (corresponding to the $\nabla^2\rho / \rho$ term in Eq.~\ref{eqn:Potential_Q}\,) 
depends on the asymptotic behavior of the density profile.
For a power-law decay $\rho(r) \propto r^{-n}$, the surface integrand scales as 
$\sqrt{\rho}\,(\nabla\sqrt{\rho}) \propto r^{-n-1}$, 
so the surface term, $\propto r^{\,1-n}$, vanishes for $n > 1$.
This condition is satisfied by all realistic NFW-like halos ($n \sim 3$ at large $r$).
Thus, in the galactic halo context 
the surface term vanishes, yielding
\begin{equation}
    \label{eqn:Q_energy_relation}
    \int \rho Q \,\mathrm{d}^3r = \frac{\hbar^2}{2m} \int |\nabla\sqrt{\rho}|^2 \,\mathrm{d}^3r \, .
\end{equation}
The prefactor $\hbar^2/(2m)$ is absorbed in the dimensionless units of Eq.~\ref{eqn:code_units}, 
so that the overall quantum energy simplifies to 
\begin{equation*}
\label{eqn:Q_energy_dimensionless}
\mathcal{Q} =  \int |\nabla\sqrt{\rho}|^2 \,\mathrm{d}^3r \, .
\end{equation*}
Hereafter, we refer to $\mathcal{Q}$ as quantum gradient energy.

Besides the quantum potential, 
another closely related term --- \emph{quantum pressure} --- is commonly used in the literature \citep{hui2017ultralight}.
It is defined as $-(1/m)\nabla Q$, as appears in the quantum Euler equation Eq.~\ref{eqn:Mad_Euler}. 
With the relations given in Eq.~\ref{eqn:Potential_Q} and Eq.~\ref{eqn:Q_energy_relation}, 
the connections among quantum pressure, the quantum potential $Q$, and the quantum gradient energy $\mathcal{Q}$ 
become straightforward.

\subsection{Generalized form of the nonlinear Schr\"odinger equation} \label{sec:methods_SPE_general}

\texttt{WaveDM.jl} supports solving the Schr\"odinger equation in the following generalized form:
\begin{equation} \label{eqn:GNLSE}
    i \frac{\partial \psi}{\partial t} = -\frac{1}{2}\nabla^2 \psi + V(\mathbf r,t,\psi) \, \psi ~,
\end{equation}
where the generic potential $V$ can be any function of position, time, and the bosonic field $\psi(\mathbf r,t)$.
Applications to wave CDM, ultracold atoms, and nonlinear optics
differ only in the specific form of $V$.
From a mathematical perspective, the potential $V$ that represents the so-called self-interactions between bosons 
can be broadly classified into two types.

(1) \textbf{Local interactions.}
In this case, the potential is a local functional of the field, depending only on the 
value of the field amplitude (equivalently the density) at the same spatial point:
\begin{equation}
    V(\mathbf r,t) = F\!\left(|\psi(\mathbf r,t)|\right) ~.
\end{equation}
A representative example is the Kerr-type nonlinearity, \mbox{$V = \kappa\,|\psi|^2$},
which is commonly used to model contact self-interactions in wave CDM and 
instantaneous nonlinear optical responses in nonlinear optics.

(2) \textbf{Nonlocal interactions.}
In this case, the potential at a given point depends on the field distribution over 
an extended spatial region. It is often written in an integral form, or equivalently 
as the solution of an auxiliary field equation such as the Poisson equation:
\begin{equation}
V(\mathbf r,t)
=
\int U(\mathbf r-\mathbf r')\,|\psi(\mathbf r',t)|^2\,d^3\mathbf r' ~ .
\end{equation}
Examples include Poisson-type potentials, such as the self-gravitational potential 
in wave CDM, as well as thermal-optical nonlinearities in nonlinear optics. 
Dipole--dipole interactions in atomic condensates provide another example 
of nonlocal interactions.

\section{Numerical Methods} \label{sec:methods_numerical}

\begin{figure*}[!tb]
    \centering
    \includegraphics[width=1.0\linewidth]{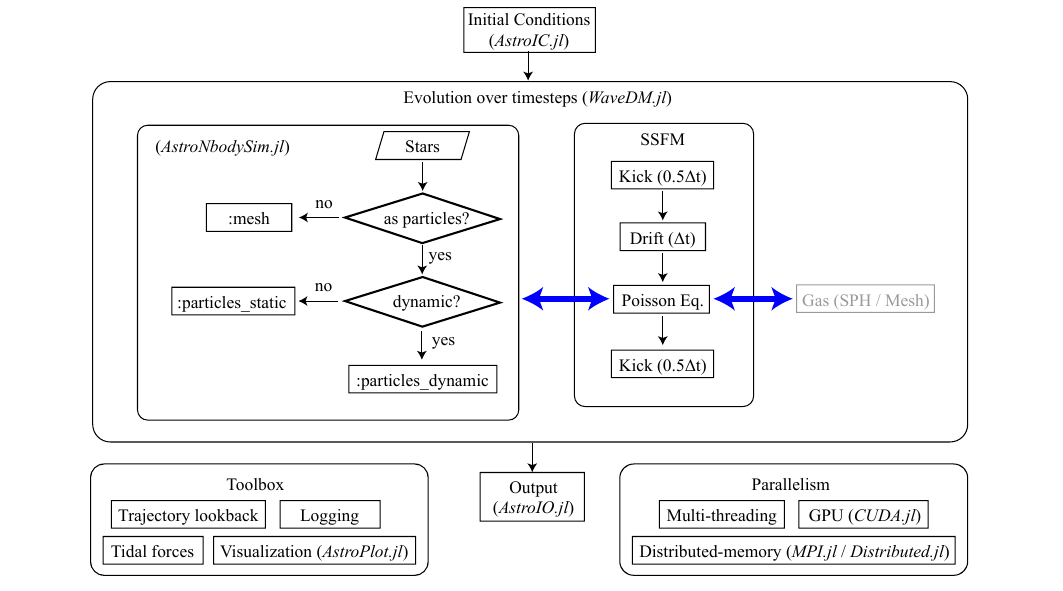}
    \caption{
        \textbf{Architecture of \texttt{WaveDM.jl}.}
        The framework integrates four core components:
        (1) a split-step Fourier method (SSFM) solver for SPE, which evolves $\psi$ using a second-order kick-drift-kick scheme;
        (2) an optional baryonic physics module (\texttt{AstroNbodySim.jl}) supporting both mesh-based and particle-based treatments, with flexible switching between static and dynamic particle modes;
        (3) a galaxy simulation toolbox providing trajectory lookback, tidal force calculations, real-time visualization, etc. (\texttt{AstroPlot.jl});
        (4) a multi-level parallel framework that combines multi-threading, distributed-memory, and GPU acceleration.
        Initial conditions are generated by \texttt{AstroIC.jl}, data I/O are handled by \texttt{AstroIO.jl}.
        Gas dynamics (SPH solver) is planned for future implementation (shown in gray).
        The thick, double-headed arrows indicate the coupling between the wave CDM and the $N$-body baryonic particles, 
        and between the wave CDM and the gas, respectively, 
        via the Poisson equation (i.e., through the total gravitational potential $\Phi_\mathrm{total}$).
    }
    \label{fig:architecture}
\end{figure*}

\texttt{WaveDM.jl} is built on a modular architecture.
As illustrated in Fig.~\ref{fig:architecture}, the core components are:
(1) a split-step Fourier method (SSFM) solver for SPE, which evolves $\psi$ using a second-order kick-drift-kick scheme;
(2) an optional baryonic physics module (\texttt{AstroNbodySim.jl})%
\footnote[3]{\url{https://github.com/JuliaAstroSim/AstroNbodySim.jl}}
supporting both particle-based ($N$-body) and mesh-based treatments, with flexible switching between static and dynamic particle modes;
(3) a galaxy simulation toolbox providing trajectory lookback, tidal force calculations, logging,
real-time visualization (\texttt{AstroPlot.jl}),%
\footnote[4]{\url{https://github.com/JuliaAstroSim/AstroPlot.jl}}
and post-processing tools;
(4) a multi-level parallel framework that combines multi-threading, distributed-memory, and GPU acceleration,
enabling the same computational workflow to scale from single-node workstations to multi-node HPC platforms.
Initial conditions are generated by \texttt{AstroIC.jl},%
\footnote[5]{\url{https://github.com/JuliaAstroSim/AstroIC.jl}}
snapshot data inputs and outputs are handled by \texttt{AstroIO.jl}.%
\footnote[6]{\url{https://github.com/JuliaAstroSim/AstroIO.jl}. All these packages are developed and maintained by R.Y.~Meng within the JuliaAstroSim ecosystem.}
Gas dynamics (smoothed particle hydrodynamics, SPH) is planned for future implementation.

\subsection{Split-step Fourier solver} \label{sec:methods_solver}

\texttt{WaveDM.jl} employs a pseudo-spectral SSFM to simulate wave dark matter dynamics.
This approach achieves spectral accuracy and computational efficiency,
making it widely adopted for nonlinear wave equations.
The fundamental strategy decomposes the evolution operator into two exactly solvable components:
\begin{equation}
    \frac{\partial \psi}{\partial t} = ({\hat{D}} + {\hat{N}}) \psi,
\end{equation}
where $\hat{D} = \frac{i}{2} \nabla^2$ represents the kinetic dispersion operator
and $\hat{N} = -i\Phi_{\mathrm{total}}$ denotes the potential operator.

We implement a second-order symmetric ``kick-drift-kick'' time-integration scheme.
Over a timestep of size $\epsilon$,
the wave function evolves according to:
\begin{equation}
    \psi(\mathbf{r}, t + \epsilon) \approx e^{\frac{\epsilon}{2} \hat{N}} e^{\epsilon \hat{D}} e^{\frac{\epsilon}{2} \hat{N}} \psi(\mathbf{r}, t).
\end{equation}
Denoting the forward Fourier transform as $\mathcal{F}$ and its inverse as $\mathcal{F}^{-1}$,
the computational sequence at each timestep is:
\begin{equation} \label{eq:SPE_solver_KDK}
    \begin{aligned}
        & \text{(a)} & \psi_1 & \leftarrow e^{\frac{\epsilon}{2} {\hat{N}}} \psi(\mathbf{r}, t) ~, \\
        & \text{(b)} & \psi_2 & \leftarrow \mathcal{F}^{-1} \left\{ e^{-\frac{i}{2} \epsilon k^2} \mathcal{F} \left\{ \psi_1 \right\} \right\} ~, \\
        & \text{(c)} & \Phi_{\mathrm{total}} & \leftarrow \mathcal{F}^{-1} \left\{ - \frac{4\pi}{k^2} \mathcal{F} \left\{ |\psi_2|^2 - \langle|\psi_2|^2\rangle \right\} \right\} + \Phi_{\mathrm{baryon}} ~, \\
        & \text{(d)} & \psi(\mathbf{r}, t+\epsilon) & \leftarrow e^{\frac{\epsilon}{2} {\hat{N}}} \psi_2 ~,
    \end{aligned}
\end{equation}
where the kinetic evolution operator $e^{-\frac{i}{2} \epsilon k^2}$ in Fourier space provides the exact solution to the dispersion term.
This sequence corresponds to:
(a) an initial half-timestep potential kick,
(b) a full kinetic drift in Fourier space,
(c) gravitational potential update from the density field,
and (d) a final half-timestep potential kick.
The Fourier transforms $\mathcal{F}$ and $\mathcal{F}^{-1}$ are implemented using the fast Fourier transform (FFT) algorithm,
which provides $\mathcal{O}(N \log N)$ complexity for grids of size $N$.
Step (c) updates the gravitational potentials,
which are computed using an FFT-based Poisson solver as described in Sect.~\ref{sec:methods_gravity}.

The simulation uses a uniform Cartesian grid with $N_x \times N_y \times N_z$ points (keywords \verb|Nx, Ny, Nz|).
For nearly spherical halos, equal grid dimensions ($N_x = N_y = N_z = N$) simplify setup,
denoted as $N^3$.
The cell sizes are defined as
\begin{equation}
    \Delta x = \frac{L_x}{N_x -1}, \Delta y = \frac{L_y}{N_y -1}, \Delta z = \frac{L_z}{N_z -1},
\end{equation}
where $L_x$, $L_y$, and $L_z$ are sidelengths of the simulation box (controlled by keywords \verb|Xmax, Ymax, Zmax|).
The coordinates of a grid point $(x_i, y_j, z_k)$ are given by
\begin{equation}
    \begin{aligned}
        x_i &= -\frac{L}{2} + i \Delta x, \quad i = 0, 1, \dots, N_x-1, \\
        y_j &= -\frac{L}{2} + j \Delta y, \quad j = 0, 1, \dots, N_y-1, \\
        z_k &= -\frac{L}{2} + k \Delta z, \quad k = 0, 1, \dots, N_z-1.
    \end{aligned}
\end{equation}
The gradient of quantities on the uniform Cartesian grid is computed using finite difference methods.
For example, at a grid point $(x_i, y_j, z_k)$,
the gradient of the order parameter $\nabla \psi$ is calculated using a second-order central differencing scheme:
{\fontsize{9.2}{11}
\begin{equation}
    \nabla \psi \approx \left( 
        \frac{\psi_{i+1,j,k} - \psi_{i-1,j,k}}{2 \Delta x},
        \frac{\psi_{i,j+1,k} - \psi_{i,j-1,k}}{2 \Delta y},
        \frac{\psi_{i,j,k+1} - \psi_{i,j,k-1}}{2 \Delta z}
    \right).
\end{equation}}

The FFT-based implementation of the SSFM
inherently assumes periodic boundary conditions for the Schr\"odinger equation.
For simulations requiring wave energy to exit the domain without reflection,
we implement a smooth absorbing layer described by the profile:
{\fontsize{9.6}{11}
\begin{equation}
    \begin{aligned}
        \mathcal{B}(\mathbf{x}) = \exp & \left[ -\alpha \left(6 - \tanh {\left(\frac{x_i + L_x/2}{L_x/N_b}\right)} + \tanh {\left(\frac{x_i - L_x/2}{L_x/N_b}\right)} \right. \right.\\
              & - \tanh\left(\frac{y_j + L_y/2}{L_y/N_b}\right) + \tanh\left(\frac{y_j - L_y/2}{L_y/N_b}\right) \\
              & \left.\left. - \tanh\left(\frac{z_k + L_z/2}{L_z/N_b}\right) + \tanh\left(\frac{z_k - L_z/2}{L_z/N_b}\right) \right) \Delta t \right] .
    \end{aligned}
\end{equation}}
Here $\alpha$ is the absorption coefficient (keyword \verb|absorb_coeff|) and $N_b$ controls the transition width (approximately $L/N_b$).
We set $\alpha = 10$ and $N_b = 50$ as default values,
corresponding to a moderate absorption strength and a transition scale of $\mathcal{O}(10^2)$ grid points.
The absorbing boundary limits unphysical interference patterns within the simulated halo.

\begin{algorithm}[H]
\small
\caption{Split-step Fourier method for the SPE}
\label{alg:SSFM}
\begin{algorithmic}[1]
\STATE \textbf{Input:} Wave function $\psi(\mathbf{r},0)$, timestep $\Delta t$, 
  total time $T_{\max}$
\STATE \textbf{Output:} Wave function $\psi(\mathbf{r},T_{\max})$
\STATE $n_{\max} \gets \lfloor T_{\max} / \Delta t \rfloor$ 
  \COMMENT{number of timesteps}
\STATE $n_{\mathrm{stride}} \gets \lfloor T_{\mathrm{snap}} / \Delta t \rfloor$ 
  \COMMENT{snapshot interval}
\FOR{$n = 0$ \TO $n_{\max} - 1$}
    \STATE \COMMENT{Step (a): First half potential kick}
    \STATE $\psi \gets \exp\!\bigl(\tfrac{\Delta t}{2}\hat{N}\bigr)\psi$, 
      where $\hat{N} = -i\Phi_{\mathrm{total}}$
    \STATE \COMMENT{Step (b): Kinetic drift in Fourier space}
    \STATE $\tilde{\psi} \gets \mathcal{F}\{\psi\}$ 
      \COMMENT{forward FFT}
    \STATE $\tilde{\psi} \gets \exp\!\bigl(-\tfrac{i}{2}\Delta t\,k^{2}\bigr)
      \tilde{\psi}$ \COMMENT{$k$: wavenumber}
    \STATE $\psi \gets \mathcal{F}^{-1}\{\tilde{\psi}\}$ 
      \COMMENT{inverse FFT}
    \STATE \COMMENT{Step (c): Update gravitational potential}
    \STATE $\rho \gets |\psi|^{2}$ \COMMENT{density field}
    \STATE $\Phi_{\mathrm{DM}} \gets \mathcal{F}^{-1}\!\bigl\{-\tfrac{4\pi}{k^{2}}
      \mathcal{F}(\rho - \langle\rho\rangle)\bigr\}$
    \STATE $\Phi_{\mathrm{total}} \gets \Phi_{\mathrm{DM}} + 
      \Phi_{\mathrm{baryon}}$
    \STATE \COMMENT{Step (d): Second half potential kick}
    \STATE $\psi \gets \exp\!\bigl(\tfrac{\Delta t}{2}\hat{N}\bigr)\psi$
    \STATE \COMMENT{Absorbing boundary condition}
    \STATE $\psi \gets \psi \odot \mathcal{B}(\mathbf{x})$ 
      \COMMENT{$\mathcal{B}$: boundary mask}
    \IF{$n \bmod n_{\mathrm{stride}} = 0$}
        \STATE save $\psi(\mathbf{r}, t = n\Delta t)$ 
          \COMMENT{output snapshot}
    \ENDIF
\ENDFOR
\RETURN $\psi(\mathbf{r}, T_{\max})$
\end{algorithmic}
\end{algorithm}


\subsection{Gravitational potential solver} \label{sec:methods_gravity}

\subsubsection{Treatment of wave dark matter} \label{sec:methods_gravity_WaveDM}

The total gravitational potential $\Phi_{\mathrm{total}} = \Phi_{\mathrm{DM}} + \Phi_{\mathrm{baryon}}$ updates automatically at each timestep as the density field evolves.
The wave dark matter contribution satisfies the Poisson equation
\begin{equation}
    \nabla^2 \Phi_{\mathrm{DM}} = 4 \pi \mathrm{G}\left(\rho - \langle\rho\rangle\right),
\end{equation}
where $\rho = |\psi|^2$ is the wave dark matter density distribution
and $\langle\rho\rangle$ denotes its spatial average over the computational domain.

\texttt{WaveDM.jl} offers two alternative boundary condition schemes for the Poisson solver.
The periodic scheme (default \verb|boundary=Periodic()|)
uses an FFT-based Poisson solver.
The potential is computed following Step (c) of Eq.~\ref{eq:SPE_solver_KDK}:
\begin{equation}
    \Phi_{\mathrm{DM}} = \mathcal{F}^{-1} \left\{ - \frac{1}{k^2} \mathcal{F} \left\{ |\psi|^2 - \langle|\psi|^2\rangle \right\} \right\}.
\end{equation}
To minimize deviations from isolated vacuum boundary conditions,
we set the simulation box size to at least eight times the halo's scale radius,
ensuring no substantial bulk mass crosses the boundaries.

Vacuum boundary conditions (\verb|boundary=Vacuum()|),
namely isolated boundary conditions \citep{duttachowdhury2021random},
use the four-step algorithm of \cite{james1977solutionpoissons} for isolated systems.
This method has a computational complexity of $\mathcal{O}(N^4)$ for a grid with $N^3$ cells,
limiting its use for high-resolution simulations.
Consequently, the vacuum scheme is primarily suitable for verification,
while periodic boundaries are recommended for production runs at high resolution.

\subsubsection{Treatment of baryonic components} \label{sec:methods_gravity_baryon}

Users can include baryonic components through four modes, controlled by \verb|baryon_mode|:
\verb|:ignored| (default), \verb|:mesh|, \verb|:particles_static|, and \verb|:particles_dynamic|.

The \verb|:mesh| mode places baryons on the same Cartesian grid as wave dark matter,
and solves the gravitational potentials of baryons using the same FFT Poisson solver.
Although straightforward to implement, this approach faces resolution constraints:
resolving stellar and gaseous disks typically requires prohibitively fine meshes ($\Delta x \lesssim 0.05 \, \mathrm{kpc}$),
making it suitable only for systems with spatially extended baryonic distributions (e.g., dwarf galaxies or early-type galaxies).

This limitation motivates the more flexible particle-based approach for modeling baryonic components at reasonable computational cost.
Particle-based modes compute $N$-body gravitational forces via two algorithms:
(1) \verb|::DirectSum| with $\mathcal{O}(N^2)$ complexity, suitable for moderate particle counts;
(2) \verb|::Tree|, a Barnes-Hut solver with $\mathcal{O}(N \log N)$ complexity following the Gadget-2 framework \citep{springel2001gadget,springel2005cosmological}.
These solvers are implemented in \texttt{AstroNbodySim.jl}.

\textbf{Static mode.} The \verb|:particles_static| mode represents baryons as 
$N$-body particles in static configuration 
providing a background gravitational potential throughout the simulation.
This mode is useful, for example, 
for modeling the gravitational influence of a non-evolving stellar distribution on wave dark matter dynamics.
The default force-softening length of $\sim 1 \, \mathrm{kpc}$ provides sufficient resolution for baryonic potentials while ensuring numerical stability.

\textbf{Dynamic mode.} The \verb|:particles_dynamic| mode allows baryons and the wave dark matter halo to co-evolve through gravitational coupling,
and thus the baryonic configuration is not fixed.
The coupling is bidirectional: at each SPE timestep, the $N$-body particles contribute their gravitational potential to the grid 
(the contribution of $\Phi_{\rm baryon}$ to $\Phi_\mathrm{total}$);
when computing $N$-body accelerations, the wave dark matter gravitational acceleration is interpolated from the grid to particle positions.
The $N$-body gravitational time integration follows the Gadget-2 scheme.

The SPE solver employs a fixed timestep $\Delta t$ determined by the Courant-Friedrichs-Lewy (CFL) condition (Sect.~\ref{sec:methods_timestep}),
whereas the $N$-body solver supports both fixed and adaptive timesteps.
Since one SPE timestep generally corresponds to multiple $N$-body timesteps,
we use the SPE time as the reference and advance the $N$-body component through multiple sub-steps within each SPE step.

\subsection{Initial conditions} \label{sec:methods_IC}

Empirically, appropriate tuning of the velocity field and total halo mass
reproduces observed density profiles after virialization.
The initial macroscopic wave function amplitude derives from the target density profile,
$|\psi| = \sqrt{\rho_{\mathrm{halo}} / m}$.
The wave function phase $\theta$ encodes the velocity field through $\mathbf{v} = \nabla \theta$.
Two principal methods exist for phase reconstruction from a velocity field:
(1) solving the Poisson equation $\nabla^2 \theta = \nabla \cdot \mathbf{v}$,
which guarantees a curl-free velocity field using homogeneous Dirichlet boundary conditions;
(2) direct integration of gradient components along independent spatial directions.
Poisson solutions enforce irrotationality,
while direct integration offers flexibility for isolated halos.
By default, \texttt{WaveDM.jl} adopts the direct integration method.

For the velocity magnitude,
we employ the local circular velocity $v_c = \sqrt{r \, \mathrm{d}\Phi_{\mathrm{total}}/\mathrm{d}r}$
as a physically motivated estimate.
The initial velocity vector is constructed as a weighted superposition
of circular rotation and random motion:
\begin{equation} \label{eq:vel_rot}
    \mathbf{v} = \zeta_{\mathrm{vel}} \left[ \zeta_{\mathrm{rot}} \, v_{c} \, \hat{\mathbf{e}}_\phi
               + (1-\zeta_{\mathrm{rot}}) \, v_{c} \, \hat{\mathbf{e}}_{\mathrm{random}} \right],
\end{equation}
where
$\zeta_{\mathrm{vel}}$ controls the velocity magnitude (keyword \verb|velocity_ratio|),
$\zeta_{\mathrm{rot}} \in [0,1]$ specifies the rotational fraction (keyword \verb|rotational_ratio|),
$\hat{\mathbf{e}}_\phi$ denotes the azimuthal unit vector,
and $\hat{\mathbf{e}}_{\mathrm{random}}$ represents a randomly oriented unit vector.

We find that
constructing a fully random velocity vector field 
(with independent random directions on each grid cell) typically leads to 
underestimated phase gradients because such fields have minimal divergence.
Consequently, insufficient kinetic energy leads to repeated halo collapses and unrealistic density distributions.
In our experience, 
these oscillations persist at higher resolutions,
and increasing the velocity magnitude $\zeta_{\mathrm{vel}}$ alone cannot correct the velocity-field setup.

To overcome this difficulty, we implement a ``pooling'' strategy (velocity field coarse-graining)
for the velocity field (by setting \verb|bulk_perturb=true| with default \verb|bulk_size=4|):
the velocity vector is periodically reset to match that of a designated reference cell.
For example, if \verb|bulk_size=4|,
then within each $4 \times 4 \times 4$ block of grid cells,
the velocity vectors are reset to match that of the block's corner cell.
Alternatively, reference cells can be placed at the box center
or randomly distributed within the block.
This technique maintains velocity coherence while avoiding numerical artifacts in the initial velocity field.

\texttt{WaveDM.jl} initializes baryons either on the mesh or as $N$-body particles.
The positions and velocities of baryons are drawn from density profiles and distribution functions.
The particle-based approaches allow flexible specification of
arbitrary density profiles and velocity distributions,
supporting multi-component systems with complex geometries
such as warped disks, stellar streams, and other irregular configurations.

\subsection{Criteria for mesh resolution} \label{sec:methods_resolution}

The setting of spatial resolution involves trade-offs between computational cost and numerical accuracy.
As demonstrated by \cite{li2019numerical}, failing to resolve the de~Broglie wavelength leads to 
incorrect evolution of the large-scale power spectrum, even when large-scale structure is well resolved.
Moreover, galaxy-scale simulations require the simulation box to encompass the entire dark matter halo,
with sufficiently low dark matter density at the boundaries,
placing strict lower limits on the box size \citep{duttachowdhury2021random}.

Given these requirements, spatial resolution is governed by two fundamental constraints.
First, the Nyquist condition $\Delta x < \lambda_{\mathrm{dB}}/2$ ensures
adequate sampling of the de~Broglie wavelength.
Second, the domain size must satisfy \mbox{$L_x \equiv N \Delta x \gg \lambda_{\mathrm{dB}}$}
to encompass all physically relevant scales.

To illustrate these constraints,
consider a boson mass $m_{22} = 1.0$
(where $m_{22} \equiv m_a / 10^{-22} \, \mathrm{eV}$)
and a typical rotation velocity $v_{\mathrm{rot}} \sim 130 \, \mathrm{km\,s^{-1}}$.
The corresponding de~Broglie wavelength is $\lambda_{\mathrm{dB}} \sim 0.77 \, \mathrm{kpc}$,
yielding resolution requirements:
(1) $\Delta x \le 0.4 \, \mathrm{kpc}$ to satisfy Nyquist sampling,
(2) $N \ge 256$ for a cubic box of side length $80 \, \mathrm{kpc}$.

We find that the maximum resolvable velocity
$v_\mathrm{max} = (\hbar/m)(\pi/\Delta x)$
(see \cite{mocz2020galaxy,may2021structure,may2023halo})
needs to exceed escape velocities.
We adopt the conservative criterion $v_\mathrm{max} \ge 2 \max\{v_\mathrm{rot}\}$
to ensure adequate velocity resolution.
Empirically, $\Delta x \lesssim \lambda_{\mathrm{dB}}/3$
ensures unbiased velocity fields,
requiring $N \ge 512$ for $m_{22} = 1.0$.

The simulation box size $L_x$ should be approximately $10\lambda_{\mathrm{dB}}$ or larger
to suppress artificial wave transmission at periodic boundaries,
which would otherwise distort the soliton core.
For smaller boson masses,
users should increase $L_x$ proportionally.
Additionally, when solving the Poisson equation on a grid,
the simulation box size should satisfy $L_x \gtrsim 8r_s$ for adequate gravitational potential resolution.

Overall, spatial resolution primarily determines the viable mass range.
We summarize five criteria to guide resolution selection:
\begin{enumerate}
    \item $\Delta x \lesssim \lambda_{\mathrm{dB}}/3$ for resolving quantum oscillations;
    \item $\Delta x \lesssim r_s/3$ for adequate density field sampling;
    \item $\Delta x \lesssim 0.3 \, \mathrm{kpc}$ for accurate inner slope fitting;
    \item $L_x \gtrsim 10\lambda_{\mathrm{dB}}$ for suppressing boundary artifacts;
    \item $L_x \gtrsim 8r_s$ for accurate gravitational potential.
\end{enumerate}
The code prioritizes criterion (1),
dynamically adjusting $L_x$ based on $m_{22}$.
Grid dimensions use powers of two for FFT efficiency.

\subsection{Criteria for timestep and simulation duration} \label{sec:methods_timestep}

The CFL condition limits the timestep $\Delta t$,
keeping phase changes below $2\pi$ to prevent numerical aliasing.
Following established conventions 
(\cite{schwabe2016simulations,mocz2017galaxy,may2021structure,may2023halo}), 
the timestep criterion is:
\begin{equation}
    \begin{aligned}
        \Delta t &\le \Delta t_{\max} \\
        &= 0.5 \min\biggl\{\frac{4}{3\pi}\frac{m}{\hbar} \Delta x^{2},\; 2\pi\frac{\hbar}{m}\frac{1}{\max|\Phi|},\; 2\pi\frac{\hbar}{m}\frac{1}{|\kappa| \max\{|\psi|^2\}}\biggr\},
    \end{aligned}
\end{equation}
where the three terms respectively constrain phase variations from:
kinetic propagation, gravitational potential, and self-interactions.

Users can enable automatic timestep selection with \verb|autoset_timestep=true|.
This computes $\Delta t =$ \mbox{$\Delta t_{\max} \times \zeta_t\,$},
where $\zeta_t$ (keyword \verb|autoset_timestep_ratio|, default 0.9) is a safety factor.

In practice,
simulations should run long enough to ensure virial equilibrium.
The ground-state oscillation period $\tau_{00} \sim \hbar/(m_a v_{\mathrm{vir}}^2)$ sets a timescale \citep{chiang2021soliton}.
For Crater~II ($v_{\mathrm{vir}} \approx 10 \, \mathrm{km\,s^{-1}}$, $m_{22} = 50.0$),
this yields $\tau_{00} \approx 3.75 \, \mathrm{Myr}$.
Long-duration simulations ($T_{\mathrm{max}} = 6 \, \mathrm{Gyr}$) demonstrate that
halos achieve virial equilibrium within approximately $4 \, \mathrm{Gyr}$
(see Fig.~\ref{fig:evolution_virial_energies} and Fig.~\ref{fig:evolution_mass_fraction_radii}).

\subsection{Criteria for three stationary states} \label{sec:methods_virialization}

At the global level, a dark matter halo is considered virialized when its kinetic and potential energies satisfy the virial relation.
On the spatially resolved scales of the internal motions within the halo, it is well known that 
three-dimensional self-gravitating systems cannot attain thermodynamic equilibrium \citep{lynden-bell1967statisticalmechanics}. 
Instead, after mean-field relaxation they become trapped in non-equilibrium quasi-stationary states (QSSs) \citep{benetti2014nonequilibrium}. 
These QSSs, which certainly are overall virialized, sustain weak (wave) turbulence inherited from the mean-field dynamics, 
such as the ``violent relaxation'' during gravitational collapse 
\citep{klessen2000gravitational,berman2012turbulence,moon2024theory}. 
By and large, they can be approximately regarded as statistically stationary, isotropic turbulence, 
and thus their global velocity field has been routinely represented as an effective velocity-dispersion pressure term \citep{chandrasekhar1953problems,klessen2000gravitational,woo2009highresolution,moon2024theory,liao2025decipheringsolitonhalo}.

Below, we introduce three criteria provided in the code for identifying (quasi-)stationary states, 
proceeding from large-scale (global) diagnostics to small-scale (fluid) ones, 
which also corresponds to a progression from the easiest stationarity 
to the most stringent one for a halo to achieve.

\textbf{Overall energy terms to diagnose virialization.}
Three key overall (i.e., volume-integrated) energy components are tracked: 
the kinetic energy $\mathcal{K}$, the gravitational potential energy $\mathcal{V}$, and the quantum gradient energy $\mathcal{Q}$. 
The overall kinetic energy $\mathcal{K}$ is defined as the total kinetic energy of the wave dark matter particles within a halo: 
\mbox{$\mathcal{K} = \frac{1}{2} \int (\rho\mathbf{v})^2 / \rho \,\mathrm{d}\mathbf{r}$}, 
which is derived from the momentum density \mbox{$\rho\mathbf{v} = (\psi^*\nabla\psi - \psi\nabla\psi^*)/(2i)$}. 
The quantum gradient energy (defined in the overall sense) is \mbox{$\mathcal{Q} = \int |\nabla\sqrt{\rho}|^2 \,\mathrm{d}\mathbf{r}$} 
(see Sect.~\ref{sec:methods_SPE_Egradient} for its relation to the quantum potential $Q$), 
and the total gravitational potential energy is \mbox{$\mathcal{V} = \int \rho\Phi \,\mathrm{d}\mathbf{r}$}.
All the three energy components oscillate during the initial relaxation phase
and reach approximately steady values after virialization,
with the total energy \mbox{$\mathcal{E} = \mathcal{K} + \mathcal{V} + \mathcal{Q}$} broadly conserved thereafter.

For a wave dark matter halo governed by the Schr\"odinger--Poisson equations,
the virial theorem states that, at virial equilibrium, $2\mathcal{K} + 2\mathcal{Q} + \mathcal{V} = 0$\,.
This virial formula can be evaluated from a single simulation snapshot,
and thus applies also to (single-epoch) observational data of realistic galaxies.

\textbf{Mass-fraction radii to diagnose radial stationarity.}
In the code we provide the radii that enclose a series of fixed mass fractions (10\%, 20\%, …, 90\%).
By definition, the spherical symmetry is assumed,
and such radii monitor the radial mass distribution. 
The evolutionary track of those radii in example numerical simulations
are illustrated in Fig.~\ref{fig:evolution_mass_fraction_radii}.

\textbf{Velocity-dispersion profiles to diagnose QSS.}
First of all, similar to the above procedure,
we can calculate the mean velocities and standard deviations (namely velocity dispersions) from the snapshots.
Spherical coordinates are routinely adopted.
To reduce numerical errors, we calculate the angle-averaged quantities,
and thus obtain the radial profiles of those mean-velocity and velocity-dispersion components:
$\langle v_r \rangle$, $\langle v_\theta \rangle$, $\langle v_\phi \rangle$,
$\sigma_r$, $\sigma_\theta$, and $\sigma_\phi$.

A straightforward use of these radial profiles is, just as discussed above, 
to monitor their temporal evolution between snapshots. 
Once a QSS is achieved, these profiles (after coarse graining to some extent) should remain unchanged.

A more powerful usage to assess QSS
is illustrated by \cite{liao2025decipheringsolitonhalo},
which relies on a Jeans-like equation involving averaged quantities
(see \cite{moon2024theory} for a detailed theoretical derivation of such 
averaged Jeans-like equations from an Euler-like fluid equation).
This approach allows the assessment to be based on a single snapshot, 
or equivalently, on observational data of a galaxy (which are essentially single‑epoch measurements).

Specifically, the Jeans-like equation that we use 
to diagnose the quasi-stationarity of both mass distribution and velocity field (i.e., QSS)
is based on radial profiles of those velocity-related quantities, as follows:
\begin{equation} \label{eq:Jeans}
\frac{\mathrm{d}(\rho \sigma_{\mathrm{r}}^{2})}{\mathrm{d} r}+2\frac{\beta(\rho \sigma_{\mathrm{r}}^{2})}{r} - 
\rho \frac{\langle v_\theta \rangle^2 + \langle v_\phi \rangle^2}{r}
=-\rho \frac{\mathrm{d}\Phi_{\mathrm{total}}}{\mathrm{d} r}
-\rho \frac{\mathrm{d}Q}{\mathrm{d} r} \,,
\end{equation}
where $\beta=1-(\sigma_{\theta}^{2} + \sigma_{\phi}^{2})/2\sigma_{\mathrm{r}}^{2}$ is the velocity-anisotropy parameter.
The two terms on the right-hand side represent the gravitational force
and the quantum-pressure force.
The three terms on the left-hand side correspond to the radial pressure-gradient term,
the anisotropic-pressure correction,
and centrifugal support from rotational streaming, respectively.

It is clear that, to arrive at Eq.~\ref{eq:Jeans},
we have assumed stationarity ($\partial_t = 0$), negligible angle-averaged radial streaming velocity ($\langle v_r \rangle = 0$),
and that the velocity field is well captured by its first- and second-order moments (mean velocity and velocity dispersion). 
Note that we have deliberately kept  
the two angle-averaged streaming velocity components perpendicular to the radial direction,
namely $\langle v_\theta \rangle$ and $\langle v_\phi \rangle$.
For realistic galaxies and halos, the two mean velocities together give the rotational velocity.

If isotropic velocity dispersion is assumed ($\beta\equiv0$), 
the radial velocity dispersion can be obtained from the above Jeans-like equation as
\begin{equation} \label{eq:Jeans_sigma}
\sigma_{\mathrm{r}}^{2}(r)=\frac{1}{\rho (r)}
\int_{r}^{\infty}\rho (r^{\prime})\,
\left(\frac{\mathrm{d}\Phi_{\mathrm{total}}}{\mathrm{d}r^{\prime}} + \frac{\mathrm{d}Q}{\mathrm{d}r^{\prime}} - \frac{\langle v_\theta \rangle^2 + \langle v_\phi \rangle^2}{r^{\prime}}\,\right)\,\mathrm{d}r^{\prime} \,.
\end{equation}
If one further assumes that all mean streaming‑velocity components can be neglected, 
then the above equation simplifies to the one employed by \cite{liao2025decipheringsolitonhalo} (see their Equation~S11).

A halo that has reached a QSS should yield a Jeans-derived $\sigma_r(r)$
that matches the directly calculated one.
This is illustrated by the cyan dotted curve in the left panel of Fig.~\ref{fig:Crater_II_velocity_field}, 
where all three directional velocity dispersions and the azimuthal rotational motion 
balance the self-gravity (as well as quantum potential) of the Crater~II satellite halo,
when the host gravitational field from the Milky Way halo is not considered. 
In contrast, as shown in the right panel of the same figure,
once the tidal gravitational field of the main halo is included
the Jeans-based diagnostic breaks down, 
indicating that the satellite halo is not in a QSS in the tidal gravitational field of the Milky Way.

During simulations, users may choose the most appropriate criterion
for assessing quasi-stationary equilibrium according to their specific scientific goals.
Below, we summarize typical use cases for each of the three criteria as a practical guide.
\begin{itemize}
	\item If the user does not require spatially resolved information about the halo, 
	the overall energy terms provide a sufficient diagnostic.
	These quantities can be computed in real time during simulations
	at modest computational cost.
	Their time evolution provides a multi-snapshot diagnostic of the system's overall stationarity.
	
	\par\hspace{1.5em}
	Particularly, the virial formula can be employed based on a single snapshot, 
	or equivalently on actual observational data of galaxies 
	that are essentially of a single epoch in terms of galactic timescales. 
	Certainly, among the above-stated three stationarities
	the state of overall virialization is the easiest for a halo to achieve.
	
	\item If the goal is to ensure that the radial structure of the halo, quantified by $\rho(r)$,
	remains steady in time --- for example, 
	when studying density profiles, radial velocity profiles, or related diagnostics --- 
	monitoring the radii enclosing fixed mass fractions (e.g., $10\%$, $20\%$, $\dots$, $90\%$) is the appropriate choice. 
	Stabilization of these radii indicates that the angle-averaged mass distribution is steady.
	These radii are obtained directly from the density field without additional dynamical modeling,
	making this criterion computationally inexpensive.
	
	\item For monitoring the relaxation of a simulated halo toward a QSS,
	the effective velocity-dispersion profile derived from the Jeans-like equation (Eq.~\ref{eq:Jeans} and Eq.~\ref{eq:Jeans_sigma})
	provides a spatially resolved and stringent diagnostic.
	As stressed above, the Jeans-like diagnostic can be applied to a single simulation snapshot.
	Because solving those equations is computationally expensive,
	the application is restricted to post-processing of simulation snapshots.
\end{itemize}

\subsection{Parallel architecture} \label{sect:parallel}

\texttt{WaveDM.jl} provides a multi-level parallel architecture that
combines shared-memory multi-threading, distributed-memory computing, and GPU acceleration within a unified framework.
This design enables the same computational workflow
to scale from single-node workstations to GPU-equipped systems
and multi-node HPC environments without requiring major code changes.
The parallelization strategy depends on the computational task:
for FFT-based operations (the SSFM solver), multi-threading or GPU is recommended;
for initial condition generation, $N$-body force calculations, and sampling, multi-threading and distributed-memory parallelism provides efficient speedup.

By default, \texttt{WaveDM.jl} leverages multi-threading for parallel execution within each process.
Users simply launch Julia with
\verb|julia -t N_threads|
to enable multi-threading
---no additional configuration required.
For FFT operations, multi-threaded FFT can be configured as:
\begin{minipage}{\linewidth}
\begin{lstlisting}[language=Julia, caption={Enable multi-threaded FFT}]
    FFTW.set_provider!("mkl")
    FFTW.set_num_threads(Threads.nthreads())
\end{lstlisting}
\end{minipage}

For distributed-memory parallelism,
\texttt{WaveDM.jl} implements collective operations
(such as \verb|scatter|, \verb|bcast|, \verb|allgather|, \verb|allreduce|)
using Julia's native \texttt{Distributed.jl} with TCP/IP transport,
rather than the third-party \texttt{MPI.jl} \citep{byrne2021mpijljulia}.
While MPI enables highly efficient distributed-memory communication,
it typically requires users to handle low-level communication details.
This choice prioritizes accessibility and interactive debugging over data communication performance:
users can add worker processes without managing low-level communication details.
For large-scale simulations requiring maximum efficiency,
future versions will support MPI as an alternative backend
without requiring changes to user code.
To enable distributed mode, set \verb|distributed_memory=true| and add worker processes:
\begin{minipage}{\linewidth}
\begin{lstlisting}[language=Julia, caption={Distributed computing with \texttt{WaveDM.jl}}]
    using Distributed
    addprocs(4)
    @everywhere using WaveDM
\end{lstlisting}
\end{minipage}

Set \verb|gpu=true| to enable GPU acceleration via the third-party Julia package \texttt{CUDA.jl} \citep{besard2019effective},
which requires a CUDA-compatible GPU and NVIDIA's CUDA toolkit.

\begin{figure}[H]
    \centering
    \includegraphics[width=1.0\linewidth]{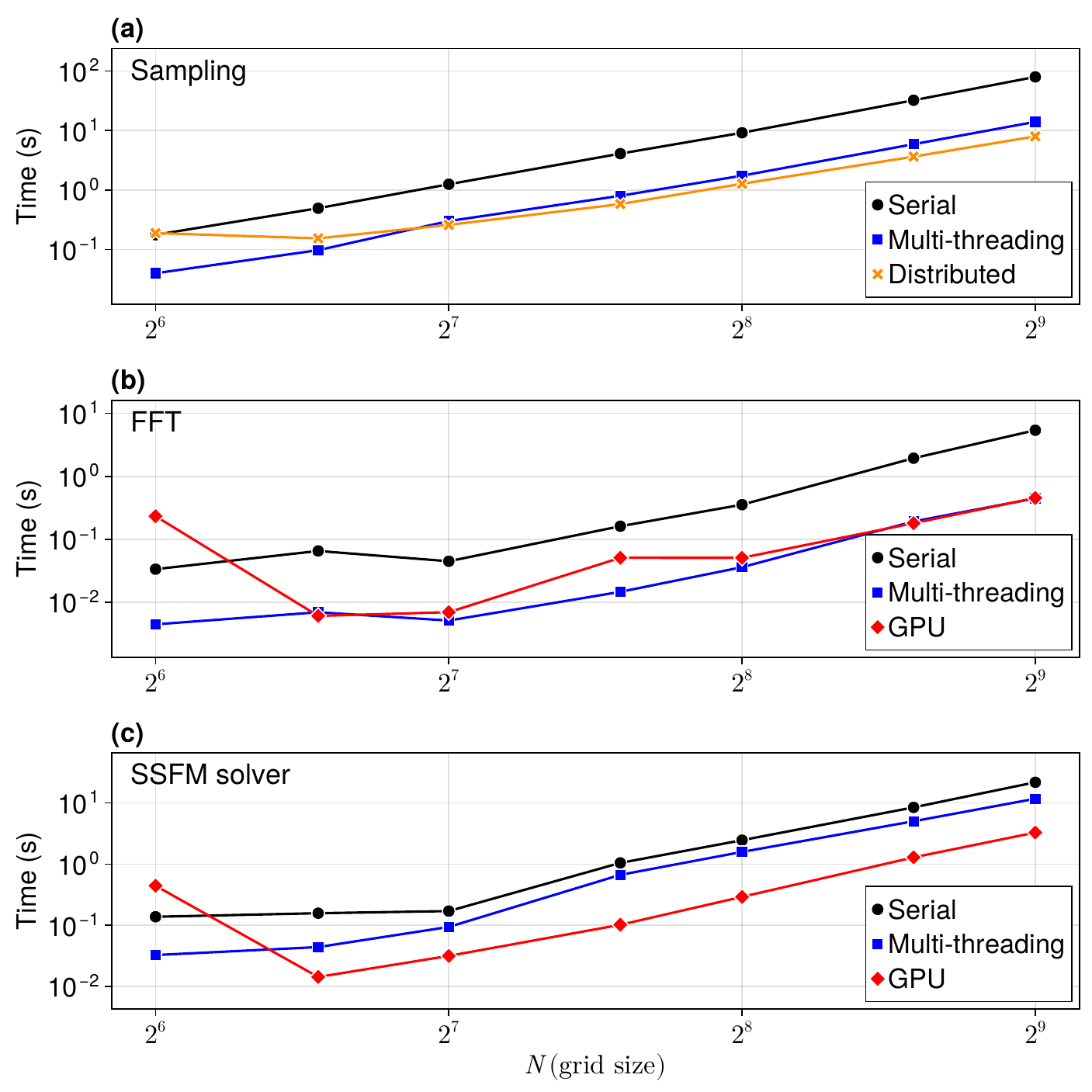}
    \caption{
        \textbf{Benchmark performance of different parallel strategies.}
        (a) Sampling. Multi-threading shows slightly lower speedup than distributed memory parallelism due to thread creation overhead;
        GPU is unsuitable due to unsupported instructions.
        (b) FFT. Multi-threading achieves 8--13$\times$ speedup;
        GPU shows no significant speedup at small $N$ due to CPU-GPU communication overhead. At large $N \gtrsim 512$, GPU becomes memory-limited (12 GB);
        distributed FFT requires MPI (planned for future).
        (c) SSFM solver. Some matrix operations are executed on GPU, achieving higher speedup than multi-threading.
    }
    \label{fig:benchmark_all}
\end{figure}

We benchmark performance on a node with two Intel Xeon E5-2698 v4 CPUs (40 cores total) and an NVIDIA Tesla P100 GPU (12 GB).
Fig.~\ref{fig:benchmark_all} shows performance for three key operations.
For sampling, multi-threading shows slightly lower speedup (6 $\times$) than distributed memory parallelism (10 $\times$) due to thread creation overhead for simple operations, while GPU acceleration is unsuitable due to unsupported instructions.
For FFT, multi-threading achieves 8--13$\times$ speedup at $N \geq 256$; due to CPU-GPU communication overhead, GPU shows no significant speedup at small $N$. At large $N \gtrsim 512$, GPU becomes memory-limited (12 GB).
For SSFM, most matrix operations are directly executed on GPU, achieving higher speedup than multi-threading; distributed FFT requires MPI support (planned for future).

\section{The Galaxy Simulation Toolbox: A Step-by-Step Tutorial with a Case Study} \label{sec:toolbox}

In this section, we provide a step-by-step tutorial on the functionalities of the galaxy-simulation toolbox within \texttt{WaveDM.jl};
this toolbox is specifically designed for studying dynamical processes of galaxy systems.
We use the Crater~II dwarf galaxy --- a satellite of the Milky Way --- as a concrete example to illustrate
the toolbox's user-friendly and comprehensive capabilities.
Further applications of \texttt{WaveDM.jl} can be found in forthcoming studies, 
including elliptical galaxies \citep{meng2026wavecold}
and dwarf galaxies (R.Y.~Meng \etal\ 2026c, in preparation).

To set up and run a simulation,
users can write a short Julia script specifying the physical parameters, initial conditions, and output options,
and then execute it from Julia's interactive terminal or from the command line interface (CLI).
As the calculation proceeds, \texttt{WaveDM.jl} can display the evolving results in interactive standalone windows,
enabling users to inspect density fields, velocity maps, and energy diagnostics in real time.

In \texttt{WaveDM.jl}, physical quantities are annotated with explicit units
(e.g., \verb|20u"kpc"|, \verb|6.0u"Gyr"|),
enabled by Julia's unit-aware arithmetic via the community-developed \texttt{Unitful.jl} package.%
\footnote[7]{\url{https://github.com/JuliaPhysics/Unitful.jl}.}
This allows dimensional consistency to be verified at compile time,
preventing common unit-conversion errors in scientific computing.
The following minimal example script simulates wave CDM dynamics for Crater~II (List~3):\\
\begin{minipage}{\linewidth}
\begin{lstlisting}[language=Julia, caption={User script: wave CDM simulation of Crater~II.}]
    ... # import necessary packages
    using WaveDM # This work

    simulate_waveDM(;
        model = :dwarf_UFDs,
        Galaxy_id = 6,         # Crater II
        V  = (x,y,z,psi)->0.0, # Potential
        Nx = 384,              # Mesh size
        Xmax = 20u"kpc",       # L_x = 2*Xmax
        Tmax = 6.0u"Gyr",
        autoset_timestep = true,
        Realtime = true,
        gpu = true,
        ... # other keywords
    )
\end{lstlisting}
\end{minipage}

\subsection{Initial condition generators}

We have developed \texttt{AstroIC.jl} for generating initial conditions with diverse density profiles.
It supports sampling arbitrary density distributions from user-defined functions,
as well as direct specification of particle properties (mass, position, and velocity) for non-equilibrium structures such as stellar streams and hypervelocity stars.
These features enable modeling of complex multi-component systems including nearby galaxies, stellar streams, and warped disks.

The DM halo of the Crater~II satellite galaxy is modeled with the generalized Navarro-Frenk-White (gNFW) profile:
\begin{equation}
    \rho_{\rm DM}(r) = \frac{\rho_0}{(r/r_s)^\beta(1 + r/r_s)^{3-\beta}},
\end{equation}
where $\rho_0$ is the characteristic density, $r_s$ the scale radius, and $\beta$ the inner slope.
Given the negligible baryonic mass fraction in Crater~II, we neglect baryonic components in its example.

The Milky Way mass model from \cite{zhu2023how} includes 
bulge, stellar disks, gas components, and dark matter halo:
\begin{equation}
    \begin{aligned}
        \rho_{\rm bulge} &= \frac{\rho_{0,b}}{(1+r'/r_0)^\alpha} \exp\left[-(r'/r_{\rm cut})^2\right], \\
        \rho_{\rm disk}(R,z) &= \frac{\Sigma_0}{2z_d} \exp\left(-|z|/z_d - R/R_d\right), \\
        \rho_{\rm gas}(R,z) &= \frac{\Sigma_0}{4z_d} \exp\left(-R_m/R - R/R_d\right) \sech^2(z/2z_d), \\
        \rho_{\rm DM}(r) &= \rho_{0,h} \left(\frac{r}{r_h}\right)^{-\gamma} \left[1 + \left(\frac{r}{r_h}\right)^\alpha\right]^{(\gamma-\beta)/\alpha},
    \end{aligned}
\end{equation}
where $r' = \sqrt{R^2 + (z/q)^2}$ in cylindrical coordinates.
The following example script illustrates sampling the Milky Way's baryons as $N$-body particles (List~4). 
These baryonic components, 
set as a static configuration in this case study,
provide the background gravitational potential used in tidal force calculations, trajectory lookback, and other analyses.
\begin{minipage}{\linewidth}
\begin{lstlisting}[language=Julia, caption={User script: sampling the Milky Way's baryons as $N$-body particles.}]
    using AstroIC

    particles_stellar_thin = generate(
        AstroIC.ExponentialDisc(;
            collection = STAR,
            NumSamples = NumSamples_thin,
            TotalMass = TotalMass_thin,
            ScaleRadius = 2.42u"kpc",
            ScaleHeight = 0.3u"kpc",
    ))

    particles_gas_HI = generate(
        AstroIC.ExponentialDisc(;
            collection = GAS,
            NumSamples = NumSamples_HI,
            TotalMass = TotalMass_HI,
            ScaleRadius = 7.0u"kpc",
            ScaleHeight = 0.085u"kpc",
            HoleRadius = 4.0u"kpc", # R_m
    ))

    # ... and other components 
\end{lstlisting}
\end{minipage}

Initial condition generators are currently available for SPARC galaxies 
(late-type, early-type, and dwarf spheroidals) and dwarf galaxies.
Additional galaxy types are continuously being integrated.

\subsection{Trajectory lookback}


\begin{figure}[H]
    \centering
    \includegraphics[width=1.0\columnwidth]{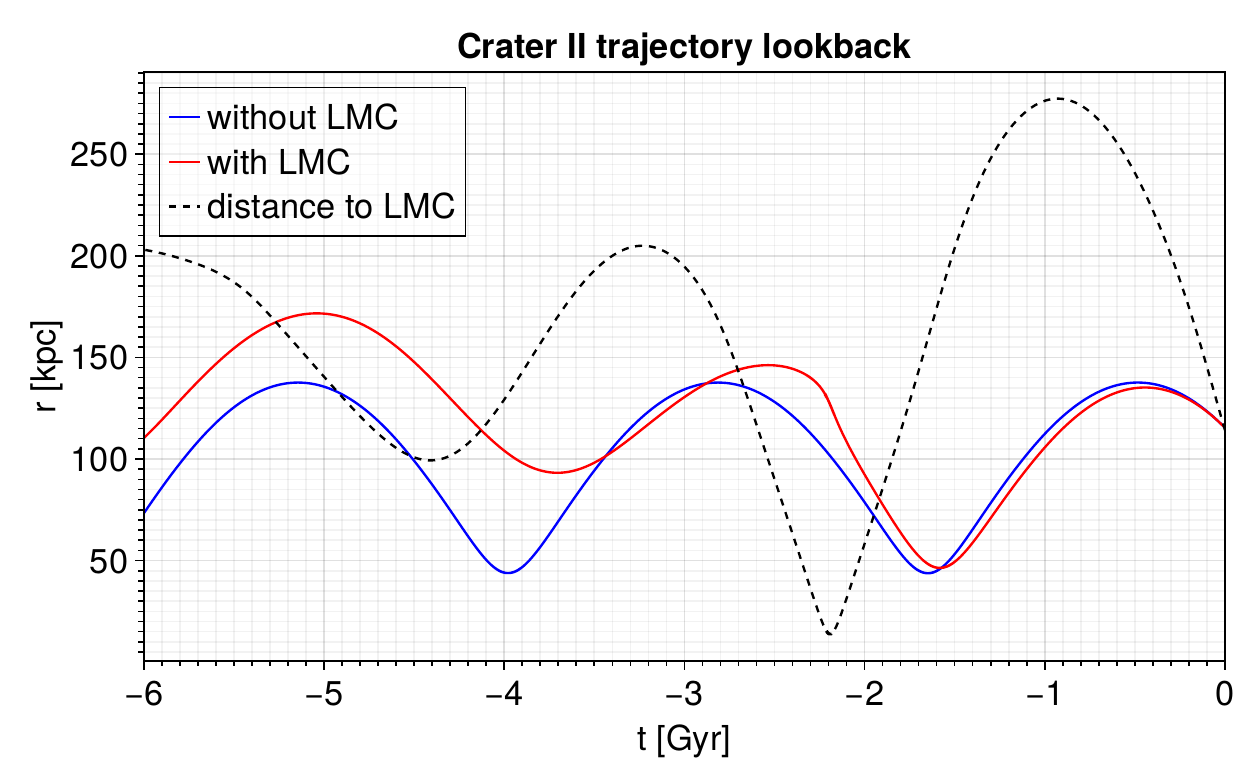}
    \caption{
        \textbf{Trajectory lookback of Crater~II.}
        Galactocentric distance as a function of lookback time.
        Blue: Milky Way only; Red: including LMC perturbations.
        Dashed line: distance to LMC.
        We find that the LMC significantly alters orbital history, particularly around $-2$ Gyr.
    }
    \label{fig:traj_Crater_II}
\end{figure}

The toolbox includes trajectory lookback for the Milky Way satellites,
based on the positional and velocity data 
listed in \cite{battaglia2022gaia}.
The tool brings those observables to the Galactocentric reference frame 
and traces orbits backward in time-dependent potentials.
Users can include LMC perturbations ($M_{\rm LMC} = 1.5\times10^{11} \, M_\odot$, $r_s = 10.84$ kpc)
by integrating the LMC orbit backward in time 
and computing its gravitational influence on the satellite galaxy of interest.
\texttt{WaveDM.jl} supports modeling LMC as either an analytic potential or 
a system of $N$-body particles.

The following demonstrates Crater~II trajectory lookback:\\
\begin{minipage}{\linewidth}
\begin{lstlisting}[language=Julia, caption={User script: trajectory lookback of Crater~II.}]
    using AstroNbodySim, WaveDM

    ## Load Crater II data
    df = AstroIC.load_data_MW_satellites()
    id = findfirst(df.Galaxy .== "Crater II")

    ## Initial conditions (Galactocentric)
    pos = PVector(df.X[id],
                  df.Y[id],
                  df.Z[id], u"kpc")
    vel = -PVector(df.v_X[id],
                   df.v_Y[id],
                   df.v_Z[id], u"km/s")

    Crater_II = StructArray([Star(uAstro; id=1)])
    Crater_II.Pos[1] = pos
    Crater_II.Vel[1] = vel

    ## N-body simulation (6 Gyr lookback)
    function MW_bg_force(particle, t)
        acc_DM = ...
        acc_b = ...
        return acc = acc_b + acc_DM
    end

    sim_traj = Simulation(deepcopy(Crater_II);
        bgforce = Function[MW_bg_force],
        TimeEnd = 6.0u"Gyr",
        TimeStep = 0.0u"Gyr", #adaptive
    );
    run(sim_traj);
\end{lstlisting}
\end{minipage}

Fig.~\ref{fig:traj_Crater_II} shows the trajectory of Crater~II in the Milky Way's gravitational potential,
with (red) and without (blue) the LMC perturbation.
The simulation results reveal that the LMC can substantially alter the satellite's orbit,
and the deviation becomes more pronounced at earlier times (especially, earlier than $\sim$ 2 Gyr ago).

\subsection{Tidal forces}

Users can include tidal forces from the Milky Way (\verb|MW_tidal_field=true|)
and perturbers like the LMC (\verb|LMC_tidal_field=true|) in their simulations.
The code computes tidal forces based on the satellite's 
instantaneous position
along its orbit.
To prevent spurious center-of-mass acceleration,
the code subtracts the spatial average of the tidal gradient (equivalently, the mean acceleration). 
Without this correction, the tidal field would exert a net force on the satellite halo, 
causing unphysical center-of-mass motion.

The dramatic effect of the Milky Way's tidal force on the satellite halo of Crater~II is clearly revealed by our simulations, 
as illustrated in Fig.~\ref{fig:Crater_II_density_slice}. 
When the tidal gravitational field of the host halo is neglected (left panel), 
the virialized Crater~II subhalo at the present day ($t = 0$) --- after 6 Gyr of evolution in the simulation ---
exhibits numerous granular structures along with the central soliton.
In contrast, once the tidal field of the host is included (right panel), 
the virialized subhalo shows almost no granules at all, 
and even the central soliton becomes significantly distorted. 
The dashed white circle in the right panel marks the tidal radius of the satellite at the final snapshot (i.e., the present epoch). 
The tidal radius $r_t$ is defined by the condition $|\vec{a}_{\mathrm{self}}(r_t)| = |\vec{a}_{\mathrm{tidal}}(r_t)|$ 
--- the radius at which the satellite's self-gravitational acceleration equals the tidal acceleration exerted by the main halo. 
Within $r_t$, the satellite remains self-gravitationally bound, 
whereas beyond $r_t$, the external tidal field dominates, and material from the subhalo can be tidally stripped.

\subsection{Logging, post-processing and visualization}

\texttt{WaveDM.jl} offers comprehensive logging and real-time visualization
for monitoring simulations and analyzing results.
The logging system tracks physical quantities including
total mass, virial energies, momentum, and on-the-fly fitting parameters.
\texttt{GLMakie.jl} \citep{danisch2021makiejlflexible} (a third-party Julia package, enabled by \verb|Realtime=true|) provides interactive 2D/3D rendering of
density fields, velocity distributions, and particle trajectories,
while \texttt{UnicodePlots.jl}~\footnote[8]{Third-party Julia package, \url{https://github.com/JuliaPlots/UnicodePlots.jl}} (\verb|unicode_plot=true|) offers terminal-based visualization
for remote environments.
Users can toggle visualization modes during execution
to inspect dynamical features such as soliton oscillations and tidal stripping.

\begin{figure*}[!htbp]
	\centering
	\includegraphics[width=1.0\textwidth]{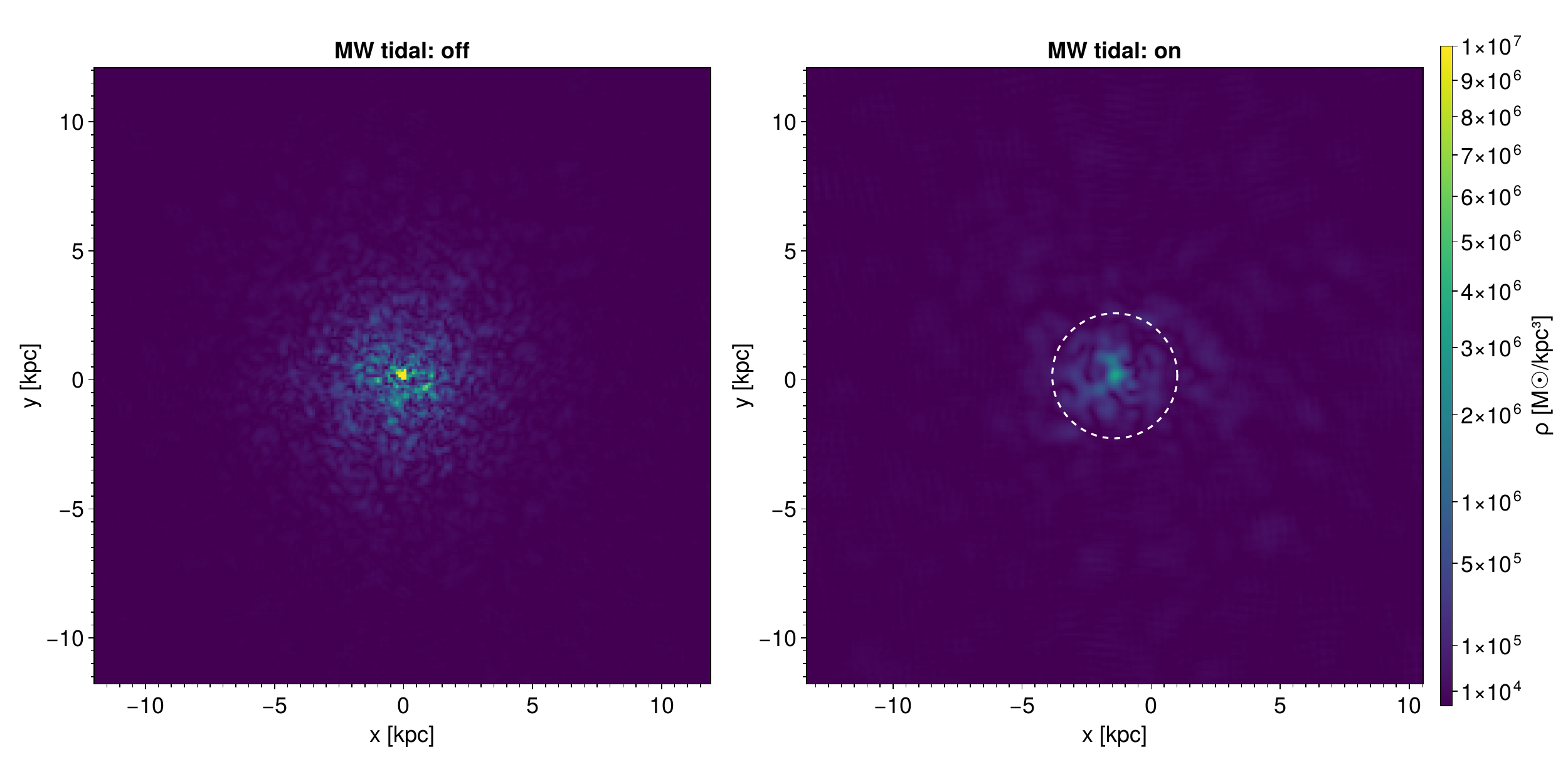}
	\caption{
		\textbf{Midplane density slice of the Crater~II satellite halo 
			at the final snapshot ($t=0$) of each simulation, without (left) and with (right) the MW tidal field.}
        Once the tidal field of the host is included, 
        the satellite halo shows almost no granules outside the tidal radius, 
        and even the central soliton becomes significantly distorted.
	}
	\label{fig:Crater_II_density_slice}
\end{figure*}

Here we demonstrate several post-processing capabilities using the Crater~II example. 
The visual impact of the tidal field has been discussed above (see \ref{fig:Crater_II_density_slice}).
The radial density profile (Fig.~\ref{fig:Crater_II_density_profile}) is consistent, within observational uncertainties, 
with the gNFW model that best fits the observational data \cite{hayashi2023dark}. 
Additional post-processing results and visualizations
--- including velocity dispersions, energy evolution, and the evolution of mass fraction radii ---
are presented in the Appendix (see Sect.~\ref{app:crater2}).

\section{Convergence tests} \label{sect:tests}

To ensure that \texttt{WaveDM.jl} produces robust results regardless of numerical choices,
we conduct systematic convergence tests.
Such validation is especially critical for Schr\"odinger-type nonlinear wave systems,
where wave interference and quantum pressure can be highly sensitive to discretization.

\begin{figure*}[!htbp]
	\centering
	\includegraphics[width=1.0\textwidth]{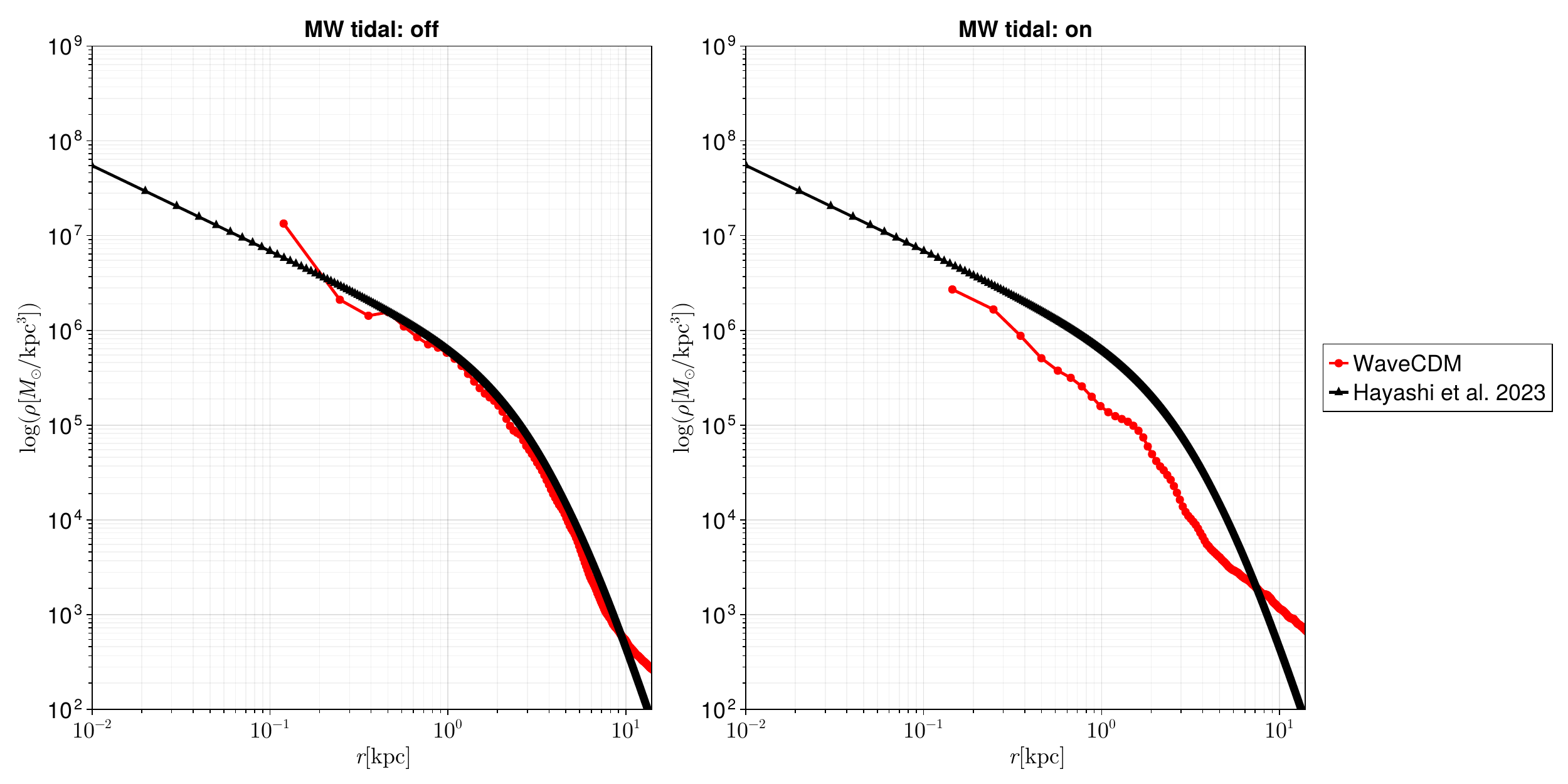}
	\caption{
		\textbf{Radial density profile of the Crater~II satellite halo
			at the final snapshot ($t=0$) of each simulation, without (left) and with (right) the MW tidal field.}
		Red curves: radially binned density profiles from our wave CDM simulations,
		computed with equally spaced radial bins of width $\Delta r \approx \Delta x$.
		Black curves:
		the gNFW profile from \cite{hayashi2023dark}.
	}
	\label{fig:Crater_II_density_profile}
\end{figure*}

The tests assess the sensitivity of the results to four key parameters: 
spatial resolution $\Delta x$, 
domain size $L$, 
timestep $\Delta t$, 
and initial condition parameters. 
The parameter ranges are summarized in Table~\ref{table:convergence_params}. 
In each test, one parameter is varied while the others are held at their baseline values. 
The baseline setup adopts the parameters used for 
the left panel of Fig.~\ref{fig:Crater_II_density_profile} (MW tidal field disabled), 
with a total simulation time of $T_{\max} = 6.0$~Gyr.

\begin{figure*}[!htbp]
	\centering
	\includegraphics[width=1.0\textwidth]{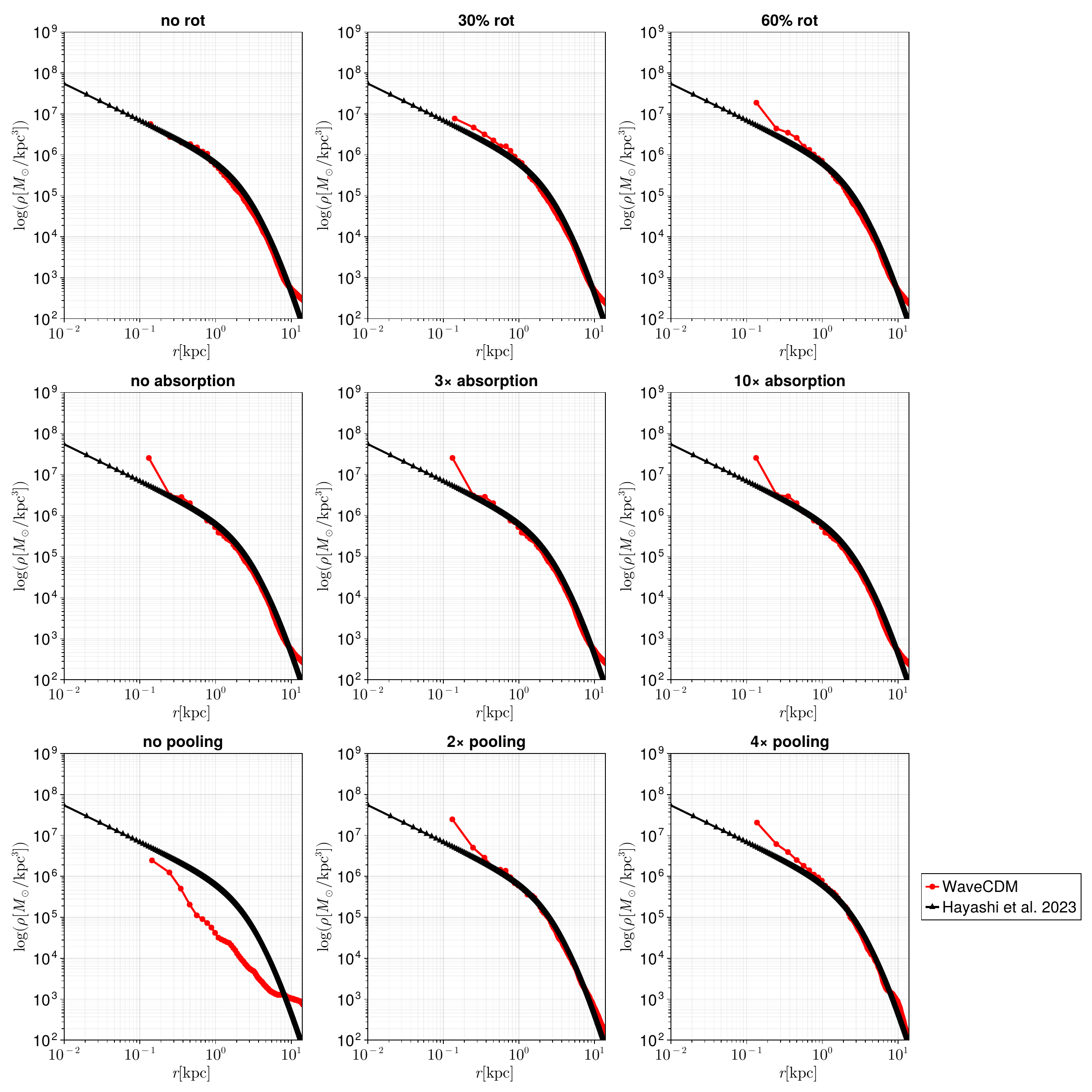}
	\caption{
    \textbf{Convergence of density profiles with respect to initial-condition parameters.}
    Each panel compares the baseline density profile (black line) 
    with a variant in which a single parameter is modified (red line). 
    The corresponding parameter is indicated in each panel title. 
    The red curves correspond to the final snapshots ($t = 0$ Gyr) of each run.
    }
	\label{fig:Convergence_profile}
\end{figure*}

The effects of varying the initial condition parameters are shown in Fig.~\ref{fig:Convergence_profile}, 
while the remaining parameters are tested in Fig.~\ref{fig:Convergence2_profile}. 
In the figures, the black curves represent the baseline density profile, 
and the red curves show the profiles obtained from the variant runs at their respective final snapshots ($t = 0$ Gyr).

\textbf{Timestep convergence.}
We find that the results are essentially insensitive to the choice of $\Delta t$. 
Varying the timestep by factors of 0.5 and 2.0 yields density and velocity dispersion profiles 
that are virtually indistinguishable from the baseline. 
This confirms that the fiducial value ($\Delta t \approx 15.9 \, \mathrm{Myr}$) 
lies well within the convergent regime, 
consistent with the stability properties of the second-order split-step integrator.

\textbf{Spatial resolution.}
Mesh spacing has a significant impact on both the initial velocity fields and the final density profiles. 
Coarser grids ($1.50\Delta x$, $N_x = 256$) tend to freeze the density field near its initial distribution.
And the phase change per timestep exceeds $2\pi$, leading to effectively randomized velocity fields. 
Finer grids ($0.75\Delta x$, $N_x = 512$), in contrast, resolve the interference patterns more accurately 
($\lambda_{\mathrm{dB}} / \Delta x \approx 8.0$), 
while leaving the overall halo structure unchanged. 
This confirms that the baseline resolution is sufficient for the present purposes.

\textbf{Domain size.}
The choice of domain size affects the outer regions of the halo primarily 
when $L$ becomes comparable to the halo scale. 
For the baseline value $L = 40 \, \mathrm{kpc}$ ($L / r_s \approx 13.3$), 
we find that boundary artifacts are negligible.

\textbf{Initial condition sensitivity.}
The velocity magnitude parameter $\zeta_{\mathrm{vel}}$ controls the initial overall kinetic energy 
and thus governs the subsequent dynamical evolution of the halo. 
We find that insufficient kinetic energy ($\zeta_{\mathrm{vel}} \lesssim 0.3$) can lead to repeated halo collapses, 
whereas excessive kinetic energy ($\zeta_{\mathrm{vel}} \gtrsim 1$) results in significant mass loss through evaporation.

Our tests further demonstrate that the ``pooling'' strategy effectively enhances the coherence of random velocity components, 
which helps produce realistic halos. 
For purely rotational velocity fields ($\zeta_{\mathrm{rot}} = 1$), 
pooling is unnecessary. 
For mixed velocity fields with both rotation and random motion ($\zeta_{\mathrm{rot}} < 1$), 
the ``pooling'' strategy provides a useful means to fine-tune the velocity field 
while preserving the overall rotational amplitude.


\begin{table*}[htbp]
	\centering
	\caption{
		\textbf{Convergence-test parameters for Crater~II.}
		The table summarizes systematic variations of six key numerical parameters around the baseline setup:
		(1) timestep $\Delta t$,
		(2) mesh spacing $\Delta x$,
		(3) simulation-box side length $L$,
		(4) absorption coefficient $\alpha$,
		(5) rotational ratio $\zeta_{\mathrm{rot}}$,
		(6) pooling parameter $\zeta_{\mathrm{pool}}$.
        The last column provides a short label for each variation.
        Note that while the parameter values in the baseline case are not bold,
        the bold values in the other cases denote the parameter values that were changed relative to the baseline case.
		The results show that $\Delta t$ has negligible impact on the final density profiles,
		whereas $\Delta x$ and $L$ significantly affect both the sampling accuracy of the density field 
		and the initial velocity-field setup.
	}
	\begin{tabular}{cccccccc}
		\hline $m_{22}$ & $N_x$          & $L$            & $N_t$        & $\zeta_{\mathrm{rot}}$ & $\alpha$     & $\zeta_{\mathrm{pool}}$     & Label \\
		&                & (kpc)          &              &               &              &             &     \\
		\hline 50       & 384            & 40             & 377          & 0.9           & 10           & 4           & baseline \\
		\hline 50       & 384            & 40             & \textbf{189} & 0.9           & 10           & 4           & 2.00 $\Delta t$ \\
		\hline 50       & 384            & 40             & \textbf{756} & 0.9           & 10           & 4           & 0.50 $\Delta t$ \\
		\hline 50       & $\mathbf{256}$ & 40             & 232          & 0.9           & 10           & 4           & 1.50 $\Delta x$ \\
		\hline 50       & $\mathbf{512}$ & 40             & 671          & 0.9           & 10           & 4           & 0.75 $\Delta x$ \\
		\hline 50       & 256            & \textbf{26.67} & 377          & 0.9           & 10           & 4           & 0.66 $L$ \\
		\hline 50       & 384            & \textbf{53.33} & 377          & 0.9           & 10           & 4           & 1.33 $L$ \\
		\hline 50       & 384            & 40             & 377          & 0.9           & \textbf{0}   & 4           & no absorb. \\
		\hline 50       & 384            & 40             & 377          & 0.9           & \textbf{30}  & 4           & 3$\times$ absorb. \\
		\hline 50       & 384            & 40             & 377          & 0.9           & \textbf{100} & 4           & 10$\times$ absorb. \\
		\hline 50       & 384            & 40             & 377          & \textbf{0.0}  & 10           & 4           & no rot \\
		\hline 50       & 384            & 40             & 377          & \textbf{0.3}  & 10           & 4           & 30\% rot \\
		\hline 50       & 384            & 40             & 377          & \textbf{0.6}  & 10           & 4           & 60\% rot \\
		\hline 50       & 384            & 40             & 377          & 0.9           & 10           & \textbf{1}  & no pooling \\
		\hline 50       & 384            & 40             & 377          & 0.9           & 10           & \textbf{8}  & 2$\times$ pooling \\
		\hline 50       & 384            & 40             & 377          & 0.9           & 10           & \textbf{16} & 4$\times$ pooling \\
		\hline
	\end{tabular}
	\label{table:convergence_params}
\end{table*}

\newcolumn

\section{Discussion}
\label{sect:discussion}

\subsection{Design philosophy: adaptability and extensibility}

\paragraph{Adaptability for general users.}
The primary design goal of \texttt{WaveDM.jl} is to make galaxy-scale simulations 
accessible to a broad range of astrophysicists,
providing them with an easy way to carry out their diverse research projects on galactic dynamics, nonequilibrium processes, and evolution.
To this end, we prioritize adaptability: the ease with which a researcher 
can reconfigure the code --- or even merely adjust a few settings --- to address their own specific scientific question.

Wave dark matter simulations often involve complex workflows
that require expertise across multiple domains:
numerical methods for the Schr\"odinger equation, parallel computing, initial conditions, $N$-body gravity, and astrophysical analysis.
\texttt{WaveDM.jl} provides an integrated framework
that unifies these components, streamlining the simulation workflow.
A typical user may wish to explore different wave CDM scenarios
by varying the boson mass $m_a$ and self-interactions,
enable or disable tidal perturbations from the Large Magellanic Cloud or similar perturbers, 
or vary the numerical resolution and initial conditions according to the problem at hand.
In all these cases, the user interacts with the code through high-level interfaces,
rather than modifying the underlying implementation.
These high-level adjustments are exposed through 
the generalised Schr\"odinger equation formulation Eq.~\ref{eqn:GNLSE}
and the pluggable potential function interface,
as illustrated in Sect.~\ref{sec:methods_SPE_general}.
For instance, users can easily switch between non-interacting and self-interacting wave CDM scenarios 
by changing the self-interaction parameter $\kappa$ in the potential definition.

The galaxy-simulation toolbox embodies this philosophy, 
providing ready-to-use tools that eliminate the need for custom code development.
\texttt{AstroIC.jl} generates initial conditions for diverse density profiles, 
and support initialization from observational kinematic data.
\texttt{AstroNbodySim.jl} enables coupled evolution of $N$-body particles and the wave dark matter field.
The trajectory lookback module reconstructs orbital histories, 
allowing users to study the dynamical evolution of satellite galaxies 
under the influence of tidal forces.
Real-time visualization provides immediate feedback during simulations, 
facilitating interactive parameter exploration.
Together, these tools lower the barrier for researchers to conduct comprehensive 
galaxy-scale studies without software engineering overhead.

\paragraph{Extensibility for advanced users.}
Beyond serving immediate users, \texttt{WaveDM.jl} is also designed as a community-oriented open-source project under the Julia ecosystem. 
For advanced users and enthusiastic programmers who wish to contribute to this code project, 
we have deliberately designed a modular architecture that facilitates extensibility.

Built upon the shared infrastructure of the JuliaAstroSim ecosystem,%
\footnote[9]{\url{https://github.com/JuliaAstroSim}, maintained by R.Y. Meng\,.}
WaveDM.jl leverages reusable components maintained by our team,
allowing developers to extend the framework without familiarity with the entire codebase:
\texttt{PhysicalParticles.jl} provides the fundamental data structures and vector algebra;
\texttt{AstroIO.jl} handles file I/O;
\texttt{AstroIC.jl} generates initial conditions;
\texttt{AstroPlot.jl} offers statistical analysis and visualization;
\texttt{PhysicalTrees.jl} 
and \texttt{PhysicalMeshes.jl} 
implement tree-based and mesh-based gravity solvers;
\texttt{PhysicalFDM.jl} 
and \texttt{PhysicalFFT.jl} 
implement FDM (finite difference method) and FFT solvers for partial differential equations (PDEs);
and \texttt{AstroNbodySim.jl} integrates all these modules and implements $N$-body methods.

New physical models --- e.g., additional baryonic feedback mechanisms --- can be added 
by implementing a small set of well-defined interfaces.
For example, SPH for gas dynamics can reuse neighbor searching algorithms from \texttt{PhysicalTrees.jl} 
in the \texttt{AstroNbodySim.jl} package.
Parallelization strategies can be extended through modifications to \texttt{PhysicalFFT.jl} and \texttt{PhysicalMeshes.jl}, 
enabling MPI-based distributed Fourier transforms for large-scale simulations.

\paragraph{} 
Julia is a modern, high-performance programming language designed specifically for scientific computing and data analysis. 
First introduced in 2009, it combines the ease of use of high-level languages with the performance of low-level ones, 
making it particularly well-suited for large-scale numerical simulations and interactive research workflows. 
These features directly support the design goals of \texttt{WaveDM.jl}: 
lowering the barrier for general users while keeping the code efficient and extensible for developers.

The code's reliance on Julia's multiple dispatch and its native distributed computing framework 
further simplifies the addition of new numerical schemes or parallelization strategies 
without disrupting existing functionality. 
By fostering an environment where contributions are accessible,
we aim to build a diverse developer community that will enrich \texttt{WaveDM.jl} with new ideas, 
maintain its long-term sustainability, and adapt it to emerging research frontiers.

\subsection{Several future developments}

We outline several planned improvements for \texttt{WaveDM.jl}, 
to be pursued either by our team or in collaboration with the community.

\textbf{Boundary conditions.}
The split-step Fourier method inherently requires periodic boundary conditions, 
which necessitates careful treatment of isolated systems using absorbing boundaries. 
Future versions will explore hybrid boundary schemes or domain-decomposition techniques to relax this constraint.

\textbf{Hydrodynamics.}
While the code supports baryonic components via $N$-body methods, 
it currently lacks hydrodynamical treatments for gas dynamics. 
We plan to incorporate a smoothed-particle hydrodynamics (SPH) solver 
for more realistic modeling of gaseous components in galactic environments.

\textbf{Cosmological simulations.}
As noted above, the current implementation focuses on galactic dynamics. 
We anticipate that the functionality for cosmological simulations will be developed 
by contributors from the Julia and astronomical communities.

\textbf{Timestep adaptation.}
The SPE solver currently uses a fixed timestep. 
We plan to implement adaptive timestepping 
based on the minimum de~Broglie wavelength across the simulation domain.

Beyond these improvements, we welcome community contributions to further enhance the code's capabilities, 
including additional initial condition generators, analysis tools, and optimization for emerging hardware architectures.

\subsection{Cross-disciplinary applications}
Beyond astrophysics, the generalized nonlinear Schr\"odinger equation framework (Eq.~\ref{eqn:GNLSE})
and the pluggable potential interface (as discussed in Sect.~\ref{sec:methods_SPE_general})
enable applications to other fields.
Specifically, \texttt{WaveDM.jl} can model nonlinear optical propagation 
and dynamical processes of ultracold atoms.
These applications share the same mathematical structure, 
making \texttt{WaveDM.jl} a general-purpose framework for various nonlinear Schr\"odinger systems,
and even enables inter-disciplinary studies among astrophysics, nonlinear optics, and condensed-matter physics.

\section{Summary}
\label{sect:summary}

We introduce \texttt{WaveDM.jl},
an open-source package for simulations of wave dark matter dynamics on galactic scales.
It is written in Julia --- 
a modern, high-performance programming language specifically designed for scientific computing and data analysis. 
The code solves the time-dependent Schr\"odinger--Poisson equation (SPE) using a second-order split-step Fourier method,
and supports simultaneous evolution of baryonic components and external perturbers through coupled $N$-body gravitational solvers.

Our primary goal is to make galaxy-scale simulations accessible to a broad range of astrophysicists, 
providing them with an easy way to carry out 
their diverse research projects without requiring deep involvement in coding and numerical implementations. 
To this end, we have deliberately designed a modular architecture 
that, by leveraging Julia's unique combination of high performance and ease of use, 
facilitates both adaptability and extensibility.

The current version of \texttt{WaveDM.jl} provides three key capabilities.
First,
it seamlessly integrates an SPE solver with $N$-body methods to simultaneously evolve wave dark matter and baryonic components,
with support for time-dependent tidal forces from host galaxies and perturbers.
Second,
its multi-level parallel architecture supports multi-threading, distributed computing, and GPU acceleration.
This design enables the same computational workflow to scale from single-node to multi-node computing environments,
without requiring major code changes from the user side.
Users can freely choose parallelization strategies in terms of their scientific purposes and available computing resources.
Third,
a dedicated galaxy-simulation toolbox provides flexible initial condition generators,
trajectory lookback, tidal force calculations, and real-time visualization capabilities.

To set up and run a simulation,
users can write a short Julia script specifying the physical parameters, initial conditions, and output options,
and then execute it from Julia's interactive terminal or from the command line interface.

Future development will expand the initial condition library,
incorporate SPH for gas dynamics,
implement adaptive timestepping,
and support MPI as an alternative communication backend.
Besides, 
leveraging its extensible design and general framework for nonlinear Schr\"odinger systems, 
\texttt{WaveDM.jl} will also be extended to implement numerical simulations 
for nonlinear optics and cold-atom applications.

The package is publicly available at \url{https://github.com/JuliaAstroSim/WaveDM.jl}
under the GNU General Public License v3.0 (GPL-3),
with comprehensive documentation and examples to support community use.

\section*{Acknowledgements}
We thank Xin-Yu Li for helpful discussions and suggestions on wave dark matter,
Zhe~Sun and Tao~Tu for years of discussions on Bose condensates and those ``super-'' phenomena,
Wen-Juan Liu for providing computational resources,
and Qian~Long for providing computational resources and guidance in Julia programming.
This work is supported by the National Natural Science Foundation of China (grant No.~12373013).
\texttt{WaveDM.jl} development benefited from the vibrant Julia community.

\section*{Conflict of interest}
The authors declare that they have no conflict of interest.

\bibliographystyle{unsrt}
\bibliography{WaveDM_code}

@article{battaglia2022gaia,
  title = {Gaia Early {{DR3}} Systemic Motions of Local Group Dwarf Galaxies and Orbital Properties with a Massive Large Magellanic Cloud},
  author = {Battaglia, G. and Taibi, S. and Thomas, G. F. and Fritz, T. K.},
  year = 2022,
  month = jan,
  journal = {Astronomy and Astrophysics},
  volume = {657},
  pages = {A54},
  issn = {0004-6361, 1432-0746},
  doi = {10.1051/0004-6361/202141528},
  urldate = {2025-07-13},
  copyright = {\copyright{} ESO 2022}
}

@article{benetti2014nonequilibrium,
  title = {Nonequilibrium Stationary States of {{3D}} Self-Gravitating Systems},
  author = {Benetti, Fernanda P. C. and {Ribeiro-Teixeira}, Ana C. and Pakter, Renato and Levin, Yan},
  year = 2014,
  month = sep,
  journal = {Physical Review Letters},
  volume = {113},
  number = {10},
  pages = {100602},
  doi = {10.1103/PhysRevLett.113.100602},
  urldate = {2026-01-06}
}

@article{berman2012turbulence,
  title = {Turbulence in a Bose-Einstein Condensate of Dipolar Excitons in Coupled Quantum Wells},
  author = {Berman, O. L. and Kezerashvili, R. {\relax Ya}. and Kolmakov, G. V. and Lozovik, {\relax Yu}. E.},
  year = 2012,
  month = jul,
  journal = {Physical Review B},
  volume = {86},
  number = {4},
  pages = {45108},
  doi = {10.1103/PhysRevB.86.045108},
  urldate = {2026-01-06}
}

@article{besard2019effective,
  title = {Effective Extensible Programming: Unleashing Julia on {{GPUs}}},
  shorttitle = {Effective Extensible Programming},
  author = {Besard, Tim and Foket, Christophe and De Sutter, Bjorn},
  year = 2019,
  month = apr,
  journal = {IEEE Transactions on Parallel and Distributed Systems},
  volume = {30},
  number = {4},
  pages = {827--841},
  issn = {1558-2183},
  doi = {10.1109/TPDS.2018.2872064},
  urldate = {2025-12-24}
}

@article{byrne2021mpijljulia,
  title = {{{MPI}}.Jl: {{Julia}} Bindings for the {{Message Passing Interface}}},
  shorttitle = {{{MPI}}.Jl},
  author = {Byrne, Simon and Wilcox, Lucas C. and Churavy, Valentin},
  year = 2021,
  month = jul,
  journal = {Proceedings of the JuliaCon Conferences},
  volume = {1},
  number = {1},
  pages = {68},
  issn = {2642-4029},
  doi = {10.21105/jcon.00068},
  urldate = {2026-03-22}
}

@article{chandrasekhar1953problems,
  title = {Problems of Gravitational Stability in the Presence of a Magnetic Field},
  author = {Chandrasekhar, S. and Fermi, E.},
  year = 1953,
  month = jul,
  journal = {The Astrophysical Journal},
  volume = {118},
  pages = {116},
  issn = {0004-637X},
  doi = {10.1086/145732},
  urldate = {2026-01-06}
}

@article{chavanis2019predictive,
  title = {Predictive Model of {{BEC}} Dark Matter Halos with a Solitonic Core and an Isothermal Atmosphere},
  author = {Chavanis, Pierre-Henri},
  year = 2019,
  journal = {Physical Review D},
  volume = {100},
  number = {8},
  pages = {083022}
}

@article{chiang2021soliton,
  title = {Soliton Oscillations and Revised Constraints from Eridanus {{II}} of Fuzzy Dark Matter},
  author = {Chiang, Barry T. and Schive, Hsi-Yu and Chiueh, Tzihong},
  year = 2021,
  journal = {Physical Review D},
  volume = {103},
  number = {10},
  pages = {103019},
  doi = {10.1103/PhysRevD.103.103019}
}

@article{danisch2021makiejlflexible,
  title = {Makie.Jl: {{Flexible}} High-Performance Data Visualization for {{Julia}}},
  shorttitle = {Makie.Jl},
  author = {Danisch, Simon and Krumbiegel, Julius},
  year = 2021,
  month = sep,
  journal = {Journal of Open Source Software},
  volume = {6},
  number = {65},
  pages = {3349},
  issn = {2475-9066},
  doi = {10.21105/joss.03349},
  urldate = {2026-03-09}
}

@misc{dong2025promise,
  title = {A Promise for the {{JWST}} Era: Massive Black Holes Directly Collapsed from Wave Dark Matter Haloes, and Star Formation in and around Their Accretion Flows},
  shorttitle = {A Promise for the {{JWST}} Era},
  author = {Dong, Xiaobo and Zhu, Yongda and Rieke, Marcia and Rieke, George and Li, Xinyu and Behroozi, Peter and Ma, Haixia and Meng, Runyu and Mao, Zhiying and Sun, Zhe},
  year = 2025,
  month = aug,
  number = {arXiv:2508.09258},
  eprint = {2508.09258},
  publisher = {arXiv},
  urldate = {2026-01-06},
  archiveprefix = {arXiv},
  howpublished = {\url{http://arxiv.org/abs/2508.09258}}
}

@article{duttachowdhury2021random,
  title = {On the {{Random Motion}} of {{Nuclear Objects}} in a {{Fuzzy Dark Matter Halo}}},
  author = {Dutta Chowdhury, Dhruba and {van den Bosch}, Frank C. and Robles, Victor H. and {van Dokkum}, Pieter and Schive, Hsi-Yu and Chiueh, Tzihong and Broadhurst, Tom},
  year = 2021,
  month = jul,
  journal = {The Astrophysical Journal},
  volume = {916},
  number = {1},
  pages = {27},
  issn = {0004-637X, 1538-4357},
  doi = {10.3847/1538-4357/ac043f},
  urldate = {2022-08-30}
}

@article{edwards2018pyultralight,
  title = {{{PyUltraLight}}: A Pseudo-Spectral Solver for Ultralight Dark Matter Dynamics},
  shorttitle = {{{PyUltraLight}}},
  author = {Edwards, Faber and Kendall, Emily and Hotchkiss, Shaun and Easther, Richard},
  year = 2018,
  month = oct,
  journal = {Journal of Cosmology and Astroparticle Physics},
  volume = {2018},
  number = {10},
  pages = {27},
  issn = {1475-7516},
  doi = {10.1088/1475-7516/2018/10/027},
  urldate = {2025-12-05}
}

@article{glennon2021modifying,
  title = {Modifying {{PyUltraLight}} to Model Scalar Dark Matter with Self-Interactions},
  author = {Glennon, Noah and {Prescod-Weinstein}, Chanda},
  year = 2021,
  month = oct,
  journal = {Physical Review D},
  volume = {104},
  number = {8},
  pages = {083532},
  doi = {10.1103/PhysRevD.104.083532},
  urldate = {2023-12-04}
}

@book{gropp1999usingmpi,
  title = {Using {{MPI}}: {{Portable Parallel Programming}} with the {{Message}}-Passing {{Interface}}},
  shorttitle = {Using {{MPI}}},
  author = {Gropp, William and Lusk, Ewing and Skjellum, Anthony},
  year = 1999,
  publisher = {MIT Press},
  isbn = {978-0-262-57132-6}
}

@article{hayashi2023dark,
  title = {Dark Matter Halo Properties of the Galactic Dwarf Satellites: Implication for Chemo-Dynamical Evolution of the Satellites and a Challenge to Lambda Cold Dark Matter},
  shorttitle = {Dark Matter Halo Properties of the Galactic Dwarf Satellites},
  author = {Hayashi, Kohei and Hirai, Yutaka and Chiba, Masashi and Ishiyama, Tomoaki},
  year = 2023,
  month = aug,
  journal = {Astrophysical Journal},
  volume = {953},
  number = {2},
  pages = {185},
  issn = {0004-637X},
  doi = {10.3847/1538-4357/ace33e},
  urldate = {2025-06-27}
}

@article{hlozek2015search,
  title = {A Search for Ultralight Axions Using Precision Cosmological Data},
  author = {Hlozek, Ren{\'e}e and Grin, Daniel and Marsh, David J. E. and Ferreira, Pedro G.},
  year = 2015,
  month = may,
  journal = {Physical Review D},
  volume = {91},
  number = {10},
  pages = {103512},
  doi = {10.1103/PhysRevD.91.103512},
  urldate = {2023-11-01}
}

@article{hopkins2019stable,
  title = {A Stable Finite-Volume Method for Scalar Field Dark Matter},
  author = {Hopkins, Philip F},
  year = 2019,
  month = oct,
  journal = {Monthly Notices of the Royal Astronomical Society},
  volume = {489},
  number = {2},
  pages = {2367--2376},
  issn = {0035-8711, 1365-2966},
  doi = {10.1093/mnras/stz1922},
  urldate = {2023-04-12}
}

@article{hu2000fuzzy,
  title = {Fuzzy {{Cold Dark Matter}}: {{The Wave Properties}} of {{Ultralight Particles}}},
  shorttitle = {Fuzzy {{Cold Dark Matter}}},
  author = {Hu, Wayne and Barkana, Rennan and Gruzinov, Andrei},
  year = 2000,
  month = aug,
  journal = {Physical Review Letters},
  volume = {85},
  number = {6},
  pages = {1158--1161},
  doi = {10.1103/PhysRevLett.85.1158},
  urldate = {2023-11-06}
}

@article{hui2017ultralight,
  title = {Ultralight Scalars as Cosmological Dark Matter},
  author = {Hui, Lam and Ostriker, Jeremiah P. and Tremaine, Scott and Witten, Edward},
  year = 2017,
  month = feb,
  journal = {Physical Review D},
  volume = {95},
  number = {4},
  pages = {043541},
  issn = {2470-0010, 2470-0029},
  doi = {10.1103/PhysRevD.95.043541},
  urldate = {2022-08-30}
}

@article{hui2021wavedark,
  title = {Wave {{Dark Matter}}},
  author = {Hui, Lam},
  year = 2021,
  month = sep,
  journal = {Annual Review of Astronomy and Astrophysics},
  volume = {59},
  number = {Volume 59, 2021},
  pages = {247--289},
  publisher = {Annual Reviews},
  issn = {0066-4146, 1545-4282},
  doi = {10.1146/annurev-astro-120920-010024},
  urldate = {2026-07-21}
}

@article{james1977solutionpoissons,
  title = {The Solution of Poisson's Equation for Isolated Source Distributions},
  author = {James, R.A},
  year = 1977,
  month = oct,
  journal = {Journal of Computational Physics},
  volume = {25},
  number = {2},
  pages = {71--93},
  issn = {00219991},
  doi = {10.1016/0021-9991(77)90013-4},
  urldate = {2026-02-25},
  copyright = {https://www.elsevier.com/tdm/userlicense/1.0/}
}

@article{kagan1997evolutioncorrelation,
  title = {Evolution of {{Correlation Properties}} and {{Appearance}} of {{Broken Symmetry}} in the {{Process}} of {{Bose-Einstein Condensation}}},
  author = {Kagan, {\relax Yu}. and Svistunov, B. V.},
  year = 1997,
  journal = {Physical Review Letters},
  volume = {79},
  number = {18},
  pages = {3331--3334},
  doi = {10.1103/PhysRevLett.79.3331}
}

@article{klessen2000gravitational,
  title = {Gravitational Collapse in Turbulent Molecular Clouds. {{I}}. {{Gasdynamical}} Turbulence},
  author = {Klessen, Ralf S. and Heitsch, Fabian and Mac Low, Mordecai-Mark},
  year = 2000,
  month = jun,
  journal = {The Astrophysical Journal},
  volume = {535},
  pages = {887--906},
  issn = {0004-637X},
  doi = {10.1086/308891},
  urldate = {2026-01-06}
}

@article{kunkel2025hybrid,
  title = {A Hybrid Scheme for Fuzzy Dark Matter Simulations Combining the Schr\"odinger and {{Hamilton}}--Jacobi--Madelung Equations},
  author = {Kunkel, Alexander and Chan, Hei Yin Jowett and Schive, Hsi-Yu and Huang, Hsinhao and Liao, Pin-Yu},
  year = 2025,
  month = jul,
  journal = {The Astrophysical Journal Supplement Series},
  volume = {279},
  number = {2},
  pages = {39},
  issn = {0067-0049},
  doi = {10.3847/1538-4365/addc59},
  urldate = {2025-11-20}
}

@article{lague2021evolving,
  title = {Evolving {{Ultralight Scalars}} into {{Non-Linearity}} with {{Lagrangian Perturbation Theory}}},
  author = {Lagu{\"e}, Alex and Bond, J. Richard and Hlo{\v z}ek, Ren{\'e}e and Marsh, David J. E. and S{\"o}ding, Laurin},
  year = 2021,
  month = apr,
  journal = {Monthly Notices of the Royal Astronomical Society},
  volume = {504},
  number = {2},
  pages = {2391--2404},
  issn = {0035-8711, 1365-2966},
  doi = {10.1093/mnras/stab601},
  urldate = {2022-08-30}
}

@article{li2019numerical,
  title = {Numerical and Perturbative Computations of the Fuzzy Dark Matter Model},
  author = {Li, Xinyu and Hui, Lam and Bryan, Greg L.},
  year = 2019,
  journal = {Physical Review D},
  volume = {99},
  number = {6},
  pages = {063509}
}

@article{liao2025decipheringsolitonhalo,
  title = {Deciphering the Soliton-Halo Relation in Fuzzy Dark Matter},
  author = {Liao, Pin-Yu and Su, Guan-Ming and Schive, Hsi-Yu and Kunkel, Alexander and Huang, Hsinhao and Chiueh, Tzihong},
  year = 2025,
  journal = {Physical Review Letters},
  volume = {135},
  number = {6},
  pages = {061002},
  doi = {10.1103/9dqj-q6mt}
}

@article{liu2025warmfuzzy,
  title = {Warm and Fuzzy Dark Matter: {{Free}} Streaming of Wave Dark Matter},
  shorttitle = {Warm and Fuzzy Dark Matter},
  author = {Liu, Rayne and Hu, Wayne and Xiao, Huangyu},
  year = 2025,
  month = jan,
  journal = {Physical Review D},
  volume = {111},
  number = {2},
  pages = {023535},
  publisher = {American Physical Society},
  doi = {10.1103/PhysRevD.111.023535},
  urldate = {2026-03-16}
}

@article{lynden-bell1967statisticalmechanics,
  title = {{{STATISTICAL MECHANICS OF VIOLENT RELAXATION IN STELLAR SYSTEMS}}},
  author = {{Lynden-Bell}, D},
  year = 1967,
  journal = {Monthly Notices of the Royal Astronomical Society},
  volume = {136},
  number = {1},
  pages = {101--121},
  issn = {0035-8711},
  doi = {10.1093/mnras/136.1.101}
}

@article{may2021structure,
  title = {Structure Formation in Large-Volume Cosmological Simulations of Fuzzy Dark Matter: {{Impact}} of the Non-Linear Dynamics},
  shorttitle = {Structure Formation in Large-Volume Cosmological Simulations of Fuzzy Dark Matter},
  author = {May, Simon and Springel, Volker},
  year = 2021,
  month = jul,
  journal = {Monthly Notices of the Royal Astronomical Society},
  volume = {506},
  number = {2},
  pages = {2603--2618},
  issn = {0035-8711, 1365-2966},
  doi = {10.1093/mnras/stab1764},
  urldate = {2022-08-31}
}

@article{may2023halo,
  title = {The Halo Mass Function and Filaments in Full Cosmological Simulations with Fuzzy Dark Matter},
  author = {May, Simon and Springel, Volker},
  year = 2023,
  month = sep,
  journal = {Monthly Notices of the Royal Astronomical Society},
  volume = {524},
  number = {3},
  pages = {4256--4274},
  issn = {0035-8711},
  doi = {10.1093/mnras/stad2031},
  urldate = {2024-12-22}
}

@article{meng2026wavecold,
  title = {Wave {{Cold Dark Matter Can Reproduce}} the {{Radial Acceleration Relation}} in {{X}}-Ray {{Elliptical Galaxies}}},
  author = {Meng, Runyu and Dong, Xiaobo},
  year = 2026,
  journal = {Research in Astronomy and Astrophysics},
  issn = {1674-4527},
  doi = {10.1088/1674-4527/aeac08},
  urldate = {2026-09-24}
}

@article{mina2020scalar,
  title = {{{SCALAR}}: An {{AMR}} Code to Simulate Axion-like Dark Matter Models},
  shorttitle = {{{SCALAR}}},
  author = {Mina, Mattia and Mota, David F. and Winther, Hans A.},
  year = 2020,
  month = sep,
  journal = {Astronomy \& Astrophysics},
  volume = {641},
  pages = {A107},
  issn = {0004-6361, 1432-0746},
  doi = {10.1051/0004-6361/201936272},
  urldate = {2023-11-01}
}

@article{mocz2015numerical,
  title = {Numerical Solution of the Nonlinear {{Schr\"odinger}} Equation Using Smoothed-Particle Hydrodynamics},
  author = {Mocz, Philip and Succi, Sauro},
  year = 2015,
  month = may,
  journal = {Physical Review E},
  volume = {91},
  number = {5},
  pages = {053304},
  issn = {1539-3755, 1550-2376},
  doi = {10.1103/PhysRevE.91.053304},
  urldate = {2023-01-25}
}

@article{mocz2017galaxy,
  title = {Galaxy Formation with {{BECDM}} -- {{I}}. {{Turbulence}} and Relaxation of Idealized Haloes},
  author = {Mocz, Philip and Vogelsberger, Mark and Robles, Victor H. and Zavala, Jes{\'u}s and {Boylan-Kolchin}, Michael and Fialkov, Anastasia and Hernquist, Lars},
  year = 2017,
  month = nov,
  journal = {Monthly Notices of the Royal Astronomical Society},
  volume = {471},
  number = {4},
  pages = {4559--4570},
  issn = {0035-8711, 1365-2966},
  doi = {10.1093/mnras/stx1887},
  urldate = {2022-08-30}
}

@article{mocz2020galaxy,
  title = {Galaxy Formation with {{BECDM}} -- {{II}}. {{Cosmic}} Filaments and First Galaxies},
  author = {Mocz, Philip and Fialkov, Anastasia and Vogelsberger, Mark and Becerra, Fernando and Shen, Xuejian and Robles, Victor H and Amin, Mustafa A and Zavala, Jes{\'u}s and {Boylan-Kolchin}, Michael and Bose, Sownak and Marinacci, Federico and Chavanis, Pierre-Henri and Lancaster, Lachlan and Hernquist, Lars},
  year = 2020,
  month = may,
  journal = {Monthly Notices of the Royal Astronomical Society},
  volume = {494},
  number = {2},
  pages = {2027--2044},
  issn = {0035-8711, 1365-2966},
  doi = {10.1093/mnras/staa738},
  urldate = {2022-08-30}
}

@misc{mocz2025jaxion,
  title = {Jaxion},
  author = {Mocz, Philip},
  year = 2025,
  month = oct,
  urldate = {2026-05-15},
  copyright = {Apache-2.0},
  howpublished = {\url{https://github.com/JaxionProject/jaxion}}
}

@article{moon2024theory,
  title = {Theory of Turbulent Equilibrium Spheres with Power-Law Linewidth--Size Relation},
  author = {Moon, Sanghyuk and Ostriker, Eve C.},
  year = 2024,
  month = nov,
  journal = {The Astrophysical Journal},
  volume = {975},
  number = {2},
  pages = {295},
  issn = {0004-637X},
  doi = {10.3847/1538-4357/ad7813},
  urldate = {2026-01-06}
}

@article{musoke2024ultradarkjl,
  title = {{{UltraDark}}.Jl: A Julia Package for Simulation of Cosmological Scalar Fields},
  shorttitle = {{{UltraDark}}.Jl},
  author = {Musoke, Nathan},
  year = 2024,
  month = apr,
  journal = {Journal of Open Source Software},
  volume = {9},
  number = {96},
  pages = {6035},
  issn = {2475-9066},
  doi = {10.21105/joss.06035},
  urldate = {2025-12-05}
}

@article{nori2018axgadget,
  title = {{{AX-GADGET}}: A New Code for Cosmological Simulations of {{Fuzzy Dark Matter}} and {{Axion}} Models},
  shorttitle = {{{AX-GADGET}}},
  author = {Nori, Matteo and Baldi, Marco},
  year = 2018,
  journal = {Monthly Notices of the Royal Astronomical Society},
  volume = {478},
  number = {3},
  pages = {3935--3951}
}

@book{pitaevskii2016boseeinstein,
  title = {Bose-Einstein Condensation and Superfluidity},
  author = {Pitaevskii, Lev and Stringari, Sandro},
  year = 2016,
  month = jan,
  publisher = {Oxford University Press},
  urldate = {2026-01-06},
  isbn = {978-0-19-875888-4}
}

@article{schive2014cosmicstructure,
  title = {Cosmic Structure as the Quantum Interference of a Coherent Dark Wave},
  author = {Schive, Hsi-Yu and Chiueh, Tzihong and Broadhurst, Tom},
  year = 2014,
  month = jul,
  journal = {Nature Physics},
  volume = {10},
  number = {7},
  pages = {496--499},
  publisher = {Nature Publishing Group},
  issn = {1745-2481},
  doi = {10.1038/nphys2996},
  urldate = {2026-05-13},
  copyright = {2014 Springer Nature Limited}
}

@article{schive2026fuzzydark,
  title = {Fuzzy Dark Matter Simulations},
  author = {Schive, Hsi-Yu},
  year = 2026,
  month = apr,
  journal = {Living Reviews in Computational Astrophysics},
  volume = {12},
  number = {1},
  pages = {1},
  issn = {2365-0524},
  doi = {10.1007/s41115-026-00027-5},
  urldate = {2026-05-08}
}

@article{schwabe2016simulations,
  title = {Simulations of Solitonic Core Mergers in Ultralight Axion Dark Matter Cosmologies},
  author = {Schwabe, Bodo and Niemeyer, Jens C. and Engels, Jan F.},
  year = 2016,
  month = aug,
  journal = {Physical Review D},
  volume = {94},
  number = {4},
  pages = {043513},
  issn = {2470-0010, 2470-0029},
  doi = {10.1103/PhysRevD.94.043513},
  urldate = {2023-01-25}
}

@article{schwabe2020simulating,
  title = {Simulating Mixed Fuzzy and Cold Dark Matter},
  author = {Schwabe, Bodo and Gosenca, Mateja and Behrens, Christoph and Niemeyer, Jens C. and Easther, Richard},
  year = 2020,
  month = oct,
  journal = {Physical Review D},
  volume = {102},
  number = {8},
  pages = {083518},
  issn = {2470-0010, 2470-0029},
  doi = {10.1103/PhysRevD.102.083518},
  urldate = {2023-11-16}
}

@article{schwabe2022deep,
  title = {Deep Zoom-in Simulation of a Fuzzy Dark Matter Galactic Halo},
  author = {Schwabe, Bodo and Niemeyer, Jens C.},
  year = 2022,
  month = may,
  journal = {Physical Review Letters},
  volume = {128},
  number = {18},
  pages = {181301},
  issn = {0031-9007, 1079-7114},
  doi = {10.1103/PhysRevLett.128.181301},
  urldate = {2022-08-30}
}

@article{springel2001gadget,
  title = {{{GADGET}}: A Code for Collisionless and Gasdynamical Cosmological Simulations},
  shorttitle = {Gadget},
  author = {Springel, Volker},
  year = 2001,
  month = apr,
  journal = {New Astronomy},
  volume = {6},
  number = {2},
  pages = {79--117},
  issn = {1384-1076},
  doi = {10.1016/S1384-1076(01)00042-2},
  urldate = {2025-04-08}
}

@article{springel2005cosmological,
  title = {The Cosmological Simulation Code {{GADGET-2}}},
  author = {Springel, Volker},
  year = 2005,
  journal = {Monthly Notices of the Royal Astronomical Society},
  volume = {364},
  number = {4},
  pages = {1105--1134},
  urldate = {2025-04-08}
}

@article{veltmaat2016cosmological,
  title = {Cosmological Particle-in-Cell Simulations with Ultralight Axion Dark Matter},
  author = {Veltmaat, Jan and Niemeyer, Jens C.},
  year = 2016,
  month = dec,
  journal = {Physical Review D},
  volume = {94},
  number = {12},
  pages = {123523},
  issn = {2470-0010, 2470-0029},
  doi = {10.1103/PhysRevD.94.123523},
  urldate = {2023-01-25}
}

@article{veltmaat2018formation,
  title = {Formation and Structure of Ultralight Bosonic Dark Matter Halos},
  author = {Veltmaat, Jan and Niemeyer, Jens C. and Schwabe, Bodo},
  year = 2018,
  month = aug,
  journal = {Physical Review D},
  volume = {98},
  number = {4},
  pages = {043509},
  issn = {2470-0010, 2470-0029},
  doi = {10.1103/PhysRevD.98.043509},
  urldate = {2023-02-14}
}

@article{woo2009highresolution,
  title = {{{HIGH-RESOLUTION SIMULATION ON STRUCTURE FORMATION WITH EXTREMELY LIGHT BOSONIC DARK MATTER}}},
  author = {Woo, Tak-Pong and Chiueh, Tzihong},
  year = 2009,
  month = may,
  journal = {The Astrophysical Journal},
  volume = {697},
  number = {1},
  pages = {850--861},
  issn = {0004-637X, 1538-4357},
  doi = {10.1088/0004-637X/697/1/850},
  urldate = {2023-10-30}
}

@article{zhang2018ultralight,
  title = {Ultralight {{Axion Dark Matter}} and {{Its Impact}} on {{Dark Halo Structure}} in {{{\emph{N}}}} -Body {{Simulations}}},
  author = {Zhang, Jiajun and Tsai, Yue-Lin Sming and Kuo, Jui-Lin and Cheung, Kingman and Chu, Ming-Chung},
  year = 2018,
  month = jan,
  journal = {The Astrophysical Journal},
  volume = {853},
  number = {1},
  pages = {51},
  issn = {1538-4357},
  doi = {10.3847/1538-4357/aaa485},
  urldate = {2022-08-30}
}

@article{zhu2023how,
  title = {How Close Dark Matter Haloes and {{MOND}} Are to Each Other: Three-Dimensional Tests Based on {{Gaia DR2}}},
  shorttitle = {How Close Dark Matter Haloes and {{MOND}} Are to Each Other},
  author = {Zhu, Yongda and Ma, Hai-Xia and Dong, Xiao-Bo and Huang, Yang and Mistele, Tobias and Peng, Bo and Long, Qian and Wang, Tianqi and Chang, Liang and Jin, Xi},
  year = 2023,
  month = mar,
  journal = {Monthly Notices of the Royal Astronomical Society},
  volume = {519},
  number = {3},
  pages = {4479--4498},
  issn = {0035-8711},
  doi = {10.1093/mnras/stac3483},
  urldate = {2023-12-08}
}

\appendix
\setcounter{section}{0}
\renewcommand{\thesection}{\Alph{section}}

\section{Appendix}

\subsection{Detailed analysis of Crater~II simulation}
\label{app:crater2}

This appendix section provides some further analyses of 
the Crater~II simulation (see Sect.~\ref{sec:toolbox}),
mainly for the purpose of illustrating the functionalities of the \texttt{WaveDM.jl} package,
particularly the different diagnostic powers of the criteria for three stationary states
described in Sect.~\ref{sec:methods_virialization}.

\subsubsection{Evolution of overall energy terms}
\begin{figure*}[!htbp]
	\centering
	\includegraphics[width=1.0\textwidth]{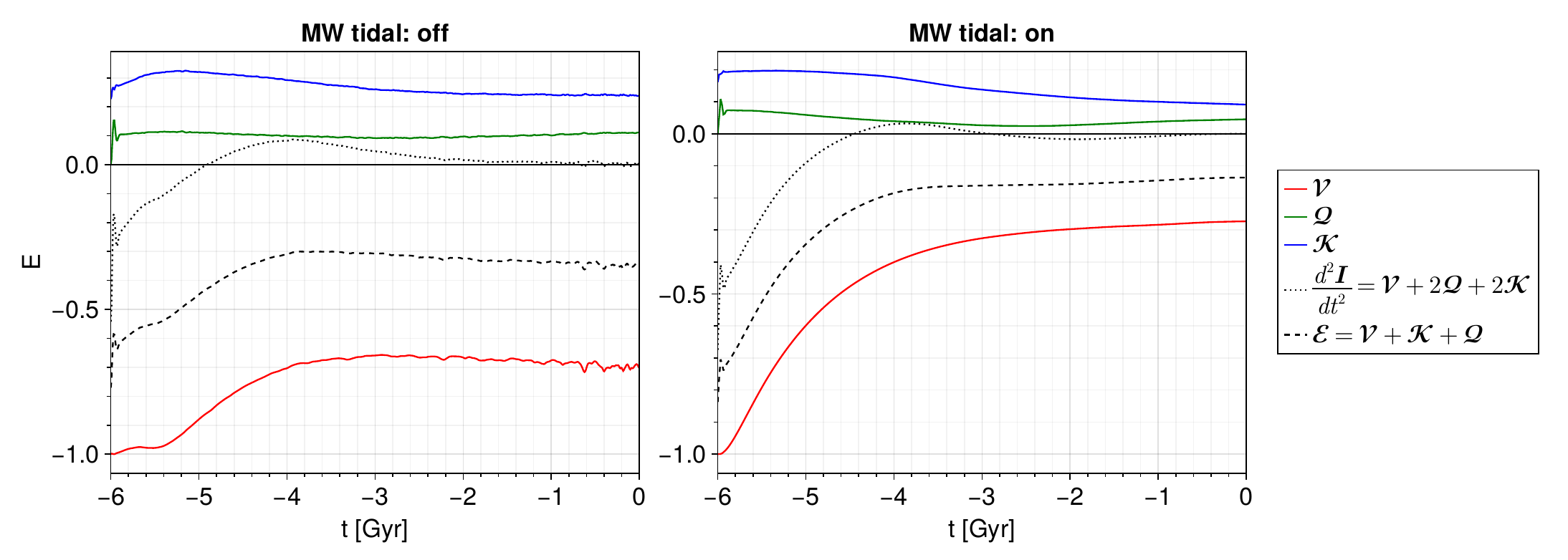}
	\caption{
		\textbf{Time evolution of the overall energy terms for the simulated Crater~II halo,
				   without (left) and with (right) the Milky Way's tidal field.}
        Time runs from $-6\,\mathrm{Gyr}$ (initial) to $0$ (present). 
   		Shown are the overall kinetic energy (blue line) $\mathcal{K} = \frac{1}{2}\int (\rho\mathbf{v})^2/\rho\,\mathrm{d}\mathbf{r}$,
		overall gravitational potential energy (red line) $\mathcal{V} = \int \rho\Phi\,\mathrm{d}\mathbf{r}$,
		quantum gradient energy (green line) $\mathcal{Q} = \int |\nabla\sqrt{\rho}|^2\,\mathrm{d}\mathbf{r}$,
		and total energy (black dashed) $\mathcal{E} = \mathcal{K} + \mathcal{V} + \mathcal{Q}$.
		In both panels, the satellite system can be virialized
		and keep the total energy broadly conserved after virialization.
        The boson mass is $m_{22}=50.0$.
	}
	\label{fig:evolution_virial_energies}
\end{figure*}

First, we illustrate the evolution of overall energy terms 
and check the virialization of the Crater~II subhalo (Fig.~\ref{fig:evolution_virial_energies}),
when the tidal gravitational field of the main halo is neglected (left panel)
and is included (right panel).

Both panels show similar evolution tracks of the overall kinetic $\mathcal{K}$, gravitational-potential energy $\mathcal{V}$,
and quantum gradient energy $\mathcal{Q}$.
In the left panel, the gravitational potential is from the subhalo only,
whereas in the right panel it is from both the subhalo and the main halo.
In both panels, starting from the initial condition ($t= -6\,\mathrm{Gyr}$), these energy terms
undergo substantial oscillations during a relaxation phase,
and then approach steady values at $t \sim -2 \,\mathrm{Gyr}$.
The total energy, $\mathcal{E}=\mathcal{V}+\mathcal{K}+\mathcal{Q}$,
is not conserved in the relaxation stage,
because the satellite halo evolves from the out-of-equilibrium initial condition and 
loses mass to some degree through the absorbing boundaries.
A similar trend is exhibited by the virial term during the relaxation stage.
But the important point to note is that
whether the tidal field of the main halo is added or not, the satellite system can be virialized
and keep the total energy conserved (broadly at least) 
after virialization.

\subsubsection{Evolution of mass-fraction radii}

\begin{figure*}[!htbp]
	\centering
	\includegraphics[width=1.0\textwidth]{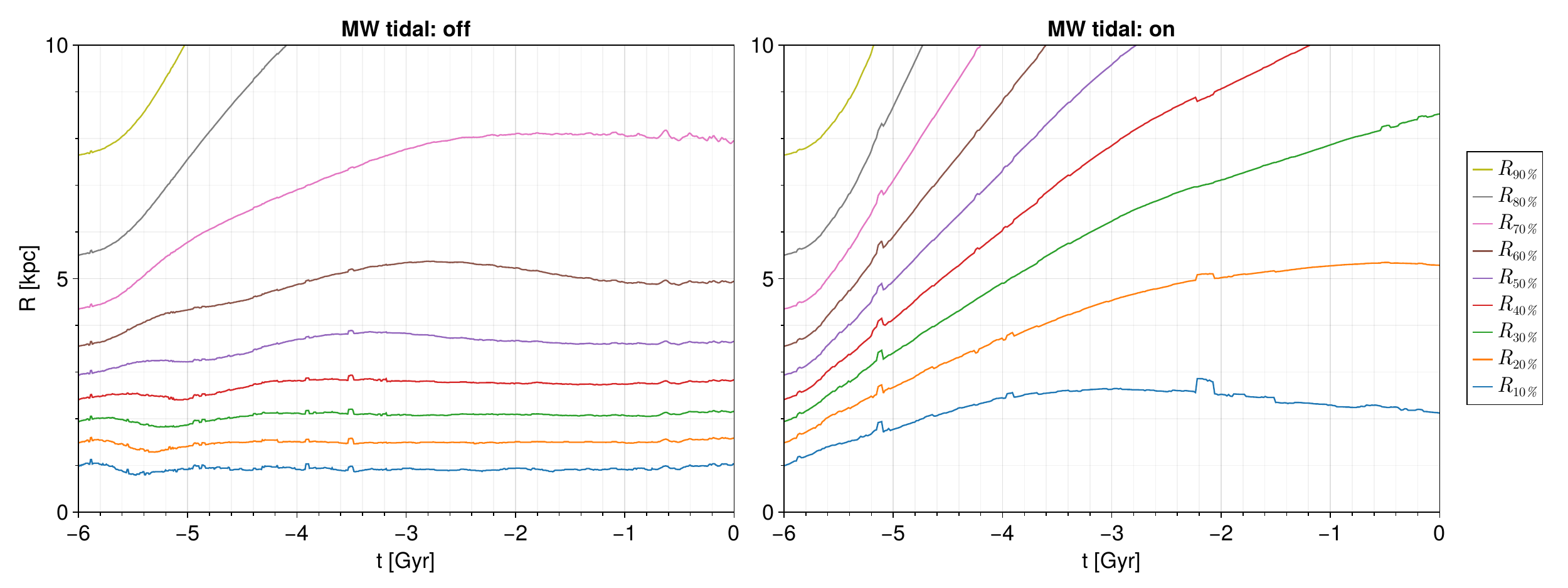}
	\caption{
       \textbf{Time evolution of mass-fraction radii for the simulated Crater~II halo, 
            without (left) and with (right) the Milky Way's tidal field.} 
        Time runs from $-6\,\mathrm{Gyr}$ (initial) to $0$ (present). 
        Without the MW field (left), the radii initially increase due to mass redistribution and 
        then settle to nearly constant values by $\sim -2\,\mathrm{Gyr}$, 
        indicating a stationary radial mass distribution. 
        With the MW field included (right), 
        the radii keep increasing throughout, sensitively signaling the non-stationary radial distribution. 
        Curves from bottom to top correspond to enclosed mass fractions of 10\%, 20\%, $\ldots$, 90\%. 
        The boson mass is $m_{22}=50.0$.
	}
	\label{fig:evolution_mass_fraction_radii}
\end{figure*}

We now illustrate the diagnostics of the stationarity of averaged radial mass distribution,
in the above-stated two cases with or without the host gravitational field.
Fig.~\ref{fig:evolution_mass_fraction_radii} shows the evolution of the radii enclosing fixed mass fractions ($10\%, 20\%, \ldots, 90\%$).
With the MW tidal field disabled (left panel), those radii initially expand due to mass redistribution during relaxation
and subsequently converge to nearly constant values, indicating the stationarity of radial mass distribution.
In contrast, with the MW tidal field enabled (right panel), 
those radii continue to evolve dramatically,
indicative of the radial non-stationarity (albeit its overall virialization).

\subsubsection{Velocity field}

\begin{figure*}[!htbp]
    \centering
    \includegraphics[width=1.0\textwidth]{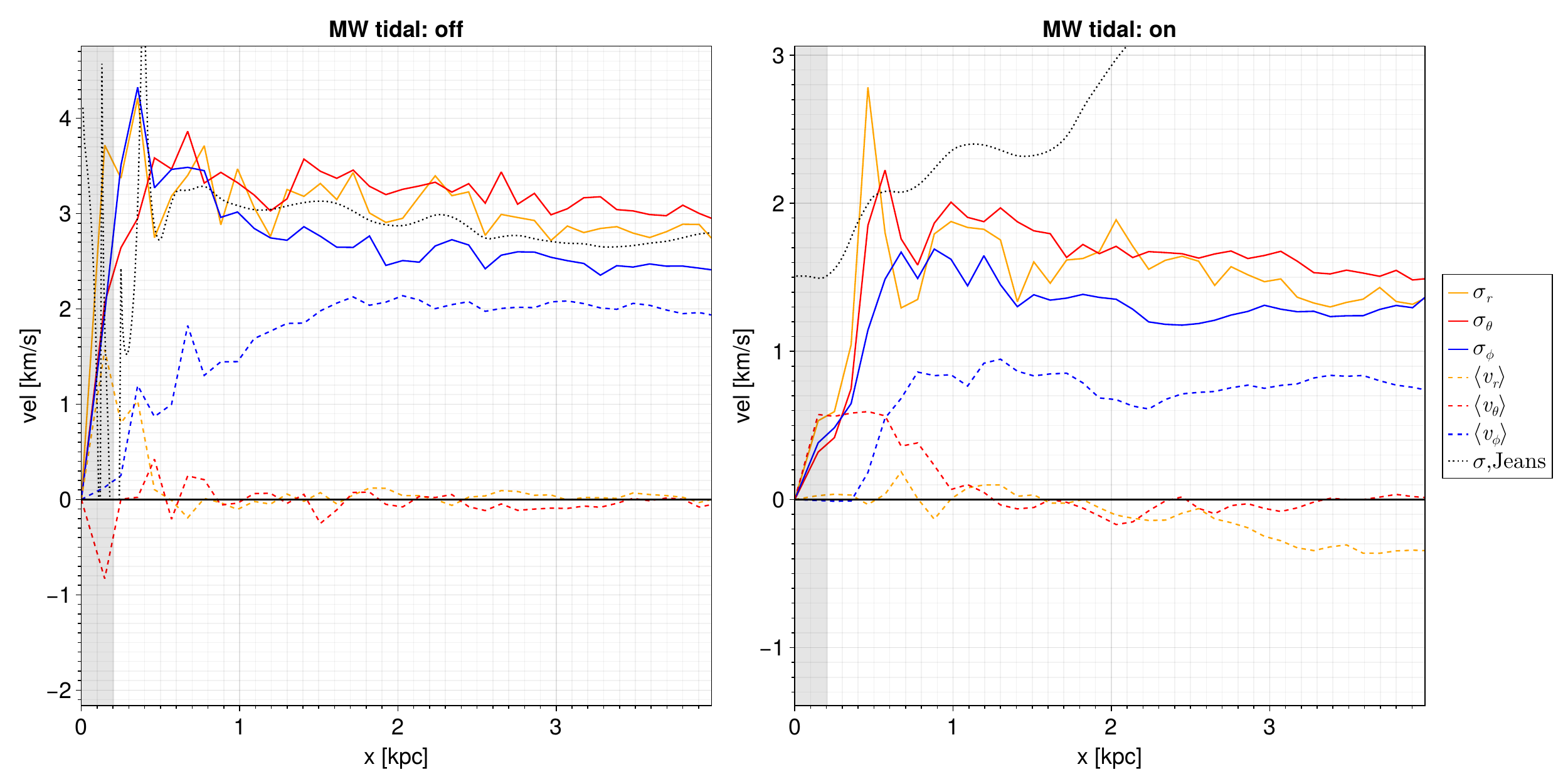}
    \caption{
        \textbf{Radial profiles of angle-averaged mean velocities and dispersions for the Crater~II halo 
        	at the final snapshot ($t=0$) of each simulation, without (left) and with (right) the MW tidal field.}
        Colors: radial (orange), zenithal (red), azimuthal (blue).
        Solid: mean velocities; Dashed: velocity dispersions;
        Dotted: Jeans prediction calculated in terms of Eq.~\ref{eq:Jeans_sigma}.
    }
    \label{fig:Crater_II_velocity_field}
\end{figure*}

Finally, we illustrate in Fig.~\ref{fig:Crater_II_velocity_field} 
the radial profiles of the velocity components after the Crater~II halo reaches overall virialization.
The velocity components are expressed in spherical coordinates:
radial ($v_r$, orange), zenithal ($v_\theta$, red), and azimuthal ($v_\phi$, blue).
Solid and dashed lines show mean velocities $\langle v_\chi \rangle$
and dispersions $\sigma_\chi$, respectively,
for each velocity component $\chi \in \{r, \theta, \phi\}$ within radial bins.
The black dotted curves show the Jeans-equation prediction $\sigma_r(r)$ from Eq.~\ref{eq:Jeans_sigma}.

\textbf{With the MW tidal field disabled (left panel),}
non-zero mean velocity in the azimuthal direction as well as
near-zero radial and zenithal mean velocities 
reflect the angular momentum inherited from the initial condition.
The radial velocity dispersion $\sigma_r(r)$ agrees with the Jeans prediction (dotted line) to within numerical scatter,
confirming that the halo has reached a QSS supported by both
velocity dispersion and the azimuthal rotational motion.

\textbf{With the MW tidal field enabled (right panel),}
tidal forces significantly alter the radial and zenithal mean velocities.
In particular, the existence of considerable $\langle v_r \rangle$
breaks the stationarity condition required by the Jeans-like equation for the angle-averaged quantities
(see Sect.~\ref{sec:methods_virialization} for details). 
Thus, the Jeans prediction of $\sigma_r(r)$ is not consistent with the measured one at all,
which just signals that the satellite halo is not in a QSS.

\subsection{Additional convergence tests}

\begin{figure*}[!htbp]
    \centering
    \includegraphics[width=1.0\textwidth]{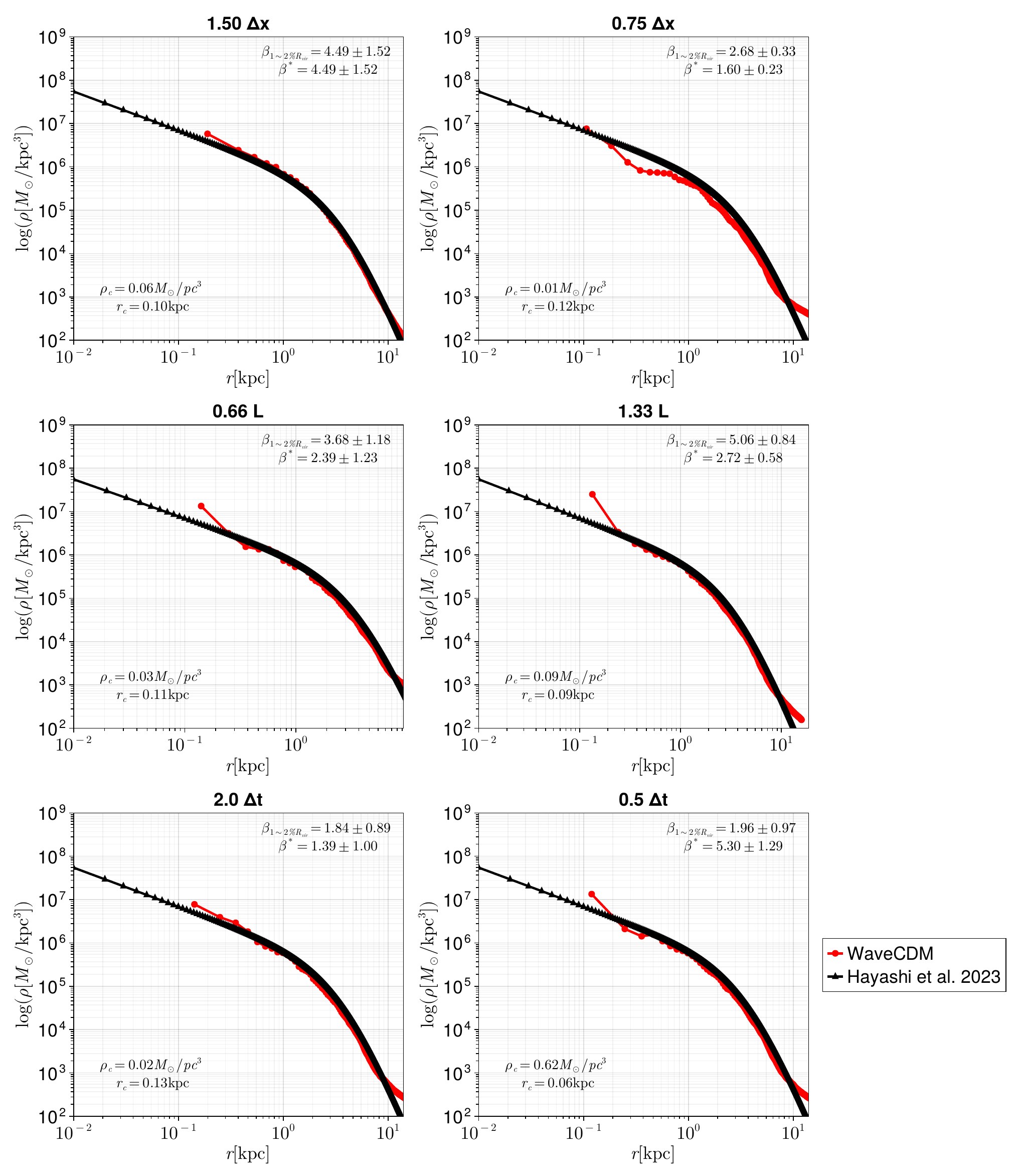}
    \caption{
        \textbf{Convergence of density profiles with respect to numerical resolution parameters.}
        Each panel compares the baseline density profile (black lines)
        with a variant in which a single resolution parameter is modified (red lines).
        The modified parameter is indicated in each panel title.
        The red lines correspond to the final snapshots ($t=0$ Gyr) of each run.
    }
    \label{fig:Convergence2_profile}
\end{figure*}

Fig.~\ref{fig:Convergence2_profile} supplements the convergence analysis of Sect.~\ref{sect:tests}
by presenting the remaining numerical resolution parameters,
including variations in mesh spacing, box size, and timestep.
 
\subsection{Existing wave CDM simulation codes}

Table~\ref{tab:codes} provides a comprehensive summary of existing wave CDM simulation codes  alongside \texttt{WaveDM.jl},
highlighting their numerical approaches, boundary condition treatments,
and cosmological capabilities.

\begin{table*}[!htbp]
    \caption{\textbf{Existing Wave CDM simulation codes}}
    \label{tab:codes}
    \centering
    \footnotesize                              
    \setlength{\tabcolsep}{4pt}
    \renewcommand{\arraystretch}{1.2}
    \resizebox{\textwidth}{!}{%
        \begin{tabular}{|c|c|c|c|c|c|c|c|c|c|}
            \hline
            Code & Reference & Solver & Boundary Condition & Gal. Dyn. with Baryons & Cosmology & Open Source \\
            \hline GAMER          & \cite{schive2014cosmicstructure,kunkel2025hybrid} & SPE, AMR      & Periodic, Isolated & $\checkmark$ & $\checkmark$ & $\checkmark$ \\ 
            \hline AxionCAMB      & \cite{hlozek2015search}                  & amended Boltzmann      & Periodic           & $\times$     & $\checkmark$ & $\checkmark$ \\ 
            \hline -              & \cite{mocz2015numerical}                 & Madelung, SPH          & Periodic           & $\times$     & $\times$ & $\times$ \\
            \hline (nyx)          & \cite{veltmaat2016cosmological}          & Madelung, PIC          & Periodic           & $\times$     & $\checkmark$ & $\times$ \\
            \hline (nyx)          & \cite{schwabe2016simulations}            & SPE, AMR               & Periodic           & $\times$     & $\times$ & $\times$ \\
            \hline (AREPO)        & \cite{mocz2017galaxy,may2021structure}   & SPE, moving-mesh       & Periodic           & $\times$     & $\times$ & $\times$ \\
            \hline (Enzo)         & \cite{veltmaat2018formation}             & hybrid, AMR            & Periodic           & ?            & $\checkmark$ & $\times$ \\
            \hline AX-GADGET      & \cite{nori2018axgadget}                  & Madelung, SPH          & Periodic           & ?            & $\checkmark$ & $\times$ \\
            \hline Axion-GADGET   & \cite{zhang2018ultralight}               & Madelung, PP           & Periodic           & ?            & $\checkmark$ & $\checkmark$ \\
            \hline PyUltraLight   & \cite{edwards2018pyultralight}           & SPE                    & Periodic           & $\times$     & $\times$ & $\checkmark$ \\
            \hline GIZMO          & \cite{hopkins2019stable}                 & Madelung, FVM          & Periodic           & ?            & $\checkmark$ & $\checkmark$ \\
            \hline SPoS (Enzo)    & \cite{li2019numerical}                   & SPE                    & Periodic           & $\checkmark$ & $\checkmark$ & $\checkmark$ \\
            \hline axionyx        & \cite{schwabe2020simulating}             & hybrid                 & Periodic           & $\times$     & $\checkmark$ & $\checkmark$ \\
            \hline SCALAR         & \cite{mina2020scalar}                    & SPE, FDM               & Periodic           & $\times$     & $\checkmark$ & $\times$ \\
            \hline AxionLPT       & \cite{lague2021evolving}                 & Madelung, Perturbation & Periodic           & $\times$     & $\checkmark$ & $\times$ \\
            \hline (axionyx)      & \cite{schwabe2022deep}                   & hybrid                 & Periodic           & $\times$     & $\checkmark$ & $\checkmark$ \\
            \hline UltraDark.jl   & \cite{musoke2024ultradarkjl}             & SPE                    & Periodic           & $\times$     & $\checkmark$ & $\checkmark$ \\
            \hline jaxion         & \cite{mocz2025jaxion}                    & SPE                    & Periodic           & $\checkmark$ & $\checkmark$ & $\checkmark$ \\
            \hline WaveDM.jl      & this work                                & SPE                    & Periodic, Isolated & $\checkmark$ & $\times$ & $\checkmark$ \\
            \hline
        \end{tabular}%
    }
\end{table*}

\end{multicols}

\end{document}